\documentclass[10pt,letterpaper,english,aps,manuscript,groupedaddress, superscriptaddress]{revtex4}
\usepackage[T1]{fontenc}
\usepackage[latin9]{inputenc}
\usepackage{color}
\usepackage{babel}
\usepackage{array}
\usepackage{multirow}
\usepackage{amsbsy}
\usepackage{amstext}
\usepackage{amssymb}
\usepackage{graphicx}
\usepackage[unicode=true,
 bookmarks=true,bookmarksnumbered=true,bookmarksopen=true,bookmarksopenlevel=1,
 breaklinks=false,pdfborder={0 0 1},backref=false,colorlinks=true]
 {hyperref}
\hypersetup{pdftitle={The LyX Tutorial},
 pdfauthor={LyX Team},
 pdfsubject={LyX-documentation Tutorial},
 pdfkeywords={LyX, documentation},
 linkcolor=blue, citecolor=blue, urlcolor=blue, filecolor=blue,pdfpagelayout=OneColumn, pdfnewwindow=true, pdfstartview=XYZ, plainpages=false}

\makeatletter

\pdfpageheight\paperheight
\pdfpagewidth\paperwidth

\providecommand{\tabularnewline}{\\}

\@ifundefined{textcolor}{}
{%
 \definecolor{BLACK}{gray}{0}
 \definecolor{WHITE}{gray}{1}
 \definecolor{RED}{rgb}{1,0,0}
 \definecolor{GREEN}{rgb}{0,1,0}
 \definecolor{BLUE}{rgb}{0,0,1}
 \definecolor{CYAN}{cmyk}{1,0,0,0}
 \definecolor{MAGENTA}{cmyk}{0,1,0,0}
 \definecolor{YELLOW}{cmyk}{0,0,1,0}
}

\usepackage{ifpdf} 
\ifpdf 

 \IfFileExists{lmodern.sty}{\usepackage{lmodern}}{}

\fi 

\let\myTOC\tableofcontents
\renewcommand\tableofcontents{%
  \frontmatter
  \pdfbookmark[1]{\contentsname}{}
  \myTOC
  \mainmatter }

\usepackage{float}
\usepackage{braket}
\usepackage{breqn}
\usepackage{multirow}

 \def\and{%
  \end{tabular}%
  \hfill%
  \hspace*{\columnsep}%
  \hfill%
  \begin{tabular}[t]{c}}%
 \def\@maketitle{%
  \newpage%
  \null%
  \vskip 2em%
  \begin{center}%
  \let \footnote \thanks%
    {\LARGE \@title \par}%
    \vskip 1.5em%
    {\large%
      \lineskip .5em%
      \hfill%
      \begin{tabular}[t]{c}%
        \@author%
      \end{tabular}\hfill\null\par}%
    \vskip 1em%
    \vskip 1em%
    {\large \@date}%
  \end{center}%
  \par%
  \vskip 1.5em}%

\makeatother

\begin{document}
\begin{center}
\textbf{\Large{Towards Tripartite Hybrid Entanglement in Quantum Dot
Molecules}}
\par\end{center}{\Large \par}

\vspace{0.5cm}

\begin{center}
\textbf{\large{M. Khoshnegar$^{1,2,3}$, A. Jafari-Salim$^{1,3}$,
M. H. Ansari$^{1,4,7}$, A. H. Majedi$^{1,3,5,6}$}}
\par\end{center}{\large \par}

\vspace{0.5cm}

\begin{center}
$^{1}$Institute for Quantum computing, Waterloo, Ontario, Canada
N2L 3G1
\par\end{center}

\vspace{-0.6cm}

\begin{center}
$^{2}$Waterloo Institute for Nanotechnology, Waterloo, Ontario, Canada
N2L 3G1
\par\end{center}

\vspace{-0.6cm}

\begin{center}
$^{3}$Department of Electrical and Computer Engineering, University
of Waterloo, Waterloo, Ontario, Canada N2L 3G1
\par\end{center}

\vspace{-0.6cm}

\begin{center}
$^{4}$Department of Physics and Astronomy, University of Waterloo,
Waterloo, Ontario, Canada N2L 3G1
\par\end{center}

\vspace{-0.6cm}

\begin{center}
$^{5}$Perimeter Institute for Theoretical Physics, Waterloo, Ontario,
Canada N2L 3G1
\par\end{center}

\vspace{-0.6cm}

\begin{center}
$^{6}$School of Engineering and Applied Sciences, Harvard University,
Cambridge, Massachusetts 02138, USA
\par\end{center}

\vspace{-0.6cm}

\begin{center}
$^{7}$Kavli Institute for Nanoscience,  Delft University of Technology, P.O.Box 5046, 2600 GA, Delft, The Netherlands
\par\end{center}

\section*{Abstract}

Establishing the hybrid entanglement among a growing number of matter
and photonic quantum bits is necessary for the scalable quantum computation and long distance quantum communication. Here we demonstrate that charged excitonic complexes forming in strongly correlated quantum dot molecules are able to generate tripartite hybrid entanglement under proper carrier quantization. The mixed orbitals of the molecule construct multilevel ground states with sub-meV hole tunneling energy and relatively large electron hybridization energy. We show that appropriate size and interdot spacing maintains the electron particle weakly localized, opening extra recombination channels by correlating ground-state excitons. This allows for creation of higher order entangled states. Nontrivial hole tunneling energy, renormalized by the multi-particle interactions, facilitates realizing the energy coincidence among only certain components of the molecule optical spectrum. This translates to the emergence of favorable spectral components in a multi-body excitonic complex which sustain principal oscillator strengths throughout the electric-field-induced hole tunneling process. We particularly analyse whether the level broadening of favorable spin configurations could be manipulated to eliminate the distinguishability of photons.

\section{Introduction}

Strong localization of charge carriers in quantum dots, maintains sufficiently long coherence time of their spin quantum bits (qubits), easing the realization of quantum correlation between the carrier spin and flying photons which is a compulsory step toward developing secure quantum communication \cite{Cirac1997}.
Experimental evidences reported so far on photon (or spin) qubit entanglement
in isolated quantum dots (QD) \cite{Benson2000,Ghali2012,Schaibley2012} are valid
manifestations of their ability in this context. Although realizing
bipartite quantum entanglement has been increasingly studied in the
recent years, there exists no theoretical proposal on generating higher
order entangled states in real quantum dot structures. In the simplest
picture, the reason lies in the anharmonic energy level spectrum of
a typical QD that impedes color matching amid its absorption energies,
making the excitonic transitions distinguishable \cite{Moreau2001,Akopian2006}.
Since the entangled degree of freedom is mostly defined based on the
polarization of photon (or spin state of carrier), this color distinguishability reduces the degree of entanglement even in a bipartite system. Current solutions offered to establish the coincidence in the energetic of photons are often perturbational \cite{Ghali2012,Gerardot2007,Stevenson2006,Khoshnegar2011,Hassler2010}
or architectural \cite{Karlsson2010,Singh2009,Khoshnegar2012}, allowing
for creating up to bipartite quantum correlated states. Actualizing
higher order entangled states, however, seems unlikely in a single
QD due to the absence of more than two color-matching excitons $X$
in its optical spectrum \cite{Persson2004}. Experimental demonstration
of tripartite entanglement has so far succeeded merely for spin states
in diamond NV centers \cite{Neumann2008} and photon states via spontaneous
parametric downconversion \cite{Shalm2012}.

Considering the fact that photon entanglement in QDs relies on the
sequential decays of correlated excitons \cite{Akopian2006}, preserving
the spatial overlaps of excitonic orbitals in the quantum dot molecule (QDM) is a requisite to maintain this correlation. Creating higher order entangled states thus demands for excitons spatially spreading across the entire QDM,
rather than a single QD, otherwise the excitons localized in different
dots are partially correlated merely via their Coulomb interactions.
This means that one constituting particle of the exciton, that is
electron or hole, has to be weakly quantized to establish a strong
correlation between the two dots. In contrast to the previous experiments
\cite{Scheibner2007,Greilich2011}, here we show that weakly-localized electron
orbitals with relatively large tunneling energy retain the correlation
and give rise to optically-active "indirect" excitonic recombinations,
opening new channels of entangled photon emission. The QDM in our
model consists two axially stacked In(Ga)As quantum dots embedded
in a GaAs nanowire \cite{Fuhrer2007}. An axial electric
field then can be exploited to drive the interdot hole tunneling,
and eventually to entangle or disentangle electron-hole pairs \cite{Bracker2006,Bester2004}.
In contrast to electron, the hole tunneling energy is noticeably renormalized
by the spin-orbit interaction in III-V materials, see Supplemental
Information and Ref. \cite{Climente2008}. Moreover, the emergence
of new entanglement channels strongly depend on the spin fine structure
of holes. Thus a detailed quantum mechanical treatment is required
to address all the underlying phenomena, including spin-orbit and
Coulomb interactions, in a few-body populated QDM.

In this paper, we study the capability of quantum dot molecules as potential sources of higher order entanglement. We show that the configurations of multiexcitons can be deterministically
controlled by tunnelling mechanism between the hybridized $s$-shells
via an external source of electric field. We analyse the particular
case of hybrid entanglement in which the spin of a carrier resting
in the metastable ground state is correlated with the polarization
of twin photons. To this end, we confine our model to the sequential
decays of a negatively (positively) charged biexciton, $XX^{-}$ ($XX^{+}$),
down to the singly charged electron (hole) state, $XX^{-}\rightarrow X^{-}\rightarrow e$ ($XX^{+}\rightarrow X^{+}\rightarrow h$), yet the scheme can be generalized to higher orders with more complexity. We demonstrate that only a limited number of transitions are favorable in the field-dependent
spectrum of QDM due to the few-body interactions and tunnelling effects. 

\section{Basic concepts and Theoretical approach}

\begin{figure}
\includegraphics[scale=0.9]{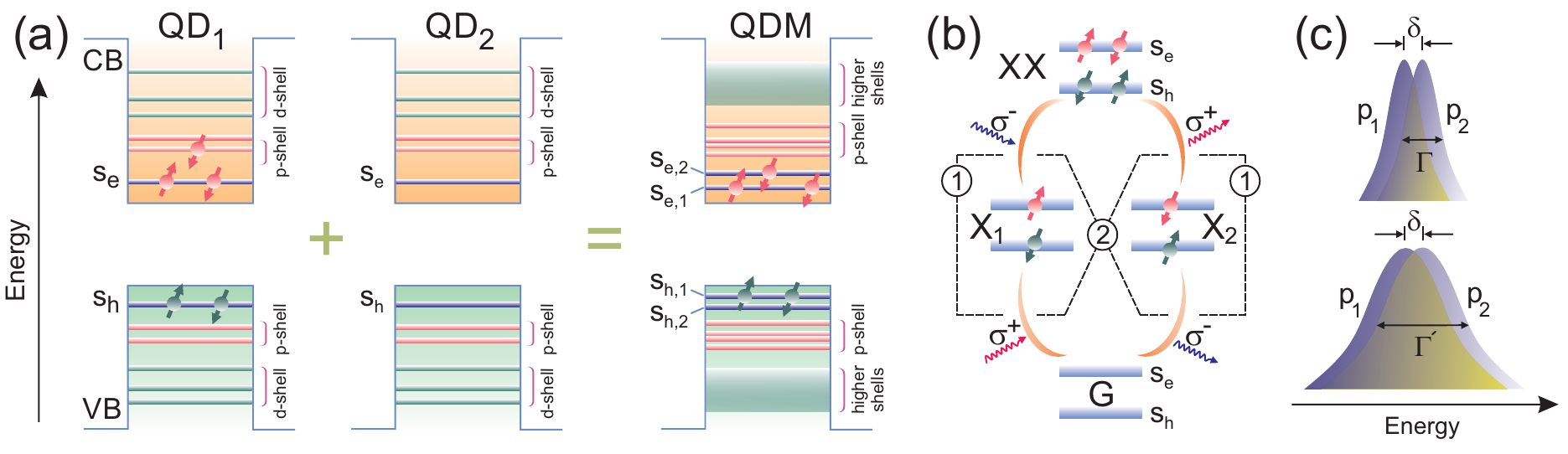}\caption{(a) Energy level spectrum of a single QD comprising $s$ ($s_{e}$,$s_{h}$),
$p$, $d$ and higher shells in both conduction and valence bands.
Once the two single dots are coupled, the energy levels shift in the
resulting QDM and their corresponding wavefunctions hybridize. One
exemplary configuration of negatively charged biexciton $XX^{-}$ is shown
in QD$_{1}$, where one electron occupies the $p$-shell. In a QDM,
however, there exists further states with $s$-like character for
the carriers to populate and form higher order complexes. (b) Bipartite
photon entanglement from a $s$-shell biexciton $XX$ in a typical
single QD: sequential decays down to the ground state $G$ give rise
to $\sigma^{+}$- and $\sigma^{-}$-polarized photons and they become
maximally entangled once the intermediate excitons $X_{1}$ and $X_{2}$
share exactly the same energy. Whether the photon pairs \textit{within
generations}, labeled by 1, or \textit{across generations} (cross-entanglement), labeled by 2, are correlated, the entanglement scheme is called regular or time reordering. (c) Spectral linewidths of partially entangled photons $P_{1}$ and $P_{2}$, separated by $\delta$ in energy, before ($\Gamma$) and after ($\Gamma^{'}$) broadening. Photons emitted within the highlighted overlap are quantum correlated.}

\label{fig:QDM shells}
\end{figure}

The energy level structure of a typical single QD is composed of $s$,
$p$, $d$ and higher shells in resemblance to the atomic states,
where the $s$-shell is referred as the ground state. In a QDM, the orbitals of each individual shell from single dots are mixed and their corresponding energy levels shift based on the coupling strength. The electron and hole $s$-shells of a single dot, $s_{e}$ and $s_{h}$, can host up to two interacting excitons
$X$ (a biexciton $XX$) thus formation of any higher order
complex, such as negatively charged biexciton $XX^{-}$ ($X+X+e$),
inevitably involves the $p$-shells; see Figure~\ref{fig:QDM shells}(a). However, $p$-shell carriers are not immune to phonon-mediated decays into the ground state, plus the initialization and quantum manipulation of carrier
spin in the $s$-shell is generally preferred. 

The QDM structure we address is a symmetric double dot embedded inside a III-V nanowire. Analogous to any other quantum confined structure, the starting point to study the QDM is to find its bound states, namely the molecular orbitals. Since we will be dealing with multi-particle states in the following sections, we refer to these molecular orbitals also  as single particle states. To solve for the molecular orbitals we consider the whole QDM as a confining heterostructure and diagonalize the modified Luttinger-Kohn hamiltonian introduced in Ref. \cite{Bahder1990}. The molecular orbitals could be approximately represented by the linear combination of dot-localized orbitals beyond the thin-barrier limit \cite{Stinaff2006}. They construct the basis set required to form the Hilbert space of multi-particle states. An electric field applied along the nanowire axis is able to change the character of molecular orbitals into the dot-localized orbitals, thus tailor the corresponding Coulomb interactions. We solve for the single and multi-particle states at several different axial electric fields and resolve the photoluminescence intensity of multi-particle transitions at grid points and midpoints by interpolation.

The QDM is designed such that the electron molecular orbitals are weakly quantized over the entire molecule and their orbitals negligibly deform in response to the axial electric field, preserving the multi-particle correlations and switching additional transitions, required for the cascade emission, on. The electron molecular orbitals remain weakly localized even after the introduction of Coulomb interactions in the multi-particle states. This is because the hole molecular orbitals are likewise delocalized at zero axial field. In contrast to the electron states, hole single particles, however, quickly localize in response to the electric field. We will demonstrate that the variation range of Coulomb interactions versus electric field is negligible compared to the electron tunneling energy.

In Sec.~\ref{sec:VI} we study the charged biexciton cascade in QDMs in the presence of hole tunneling. The emergence of favorable cascade transitions depends on the spin-fine structure of the contributing complexes, namely $XX^{-}$ and $X^{-}$ ($XX^{+}$ and $X^{+}$), and localization of the molecular orbitals under the variable electric field. Furthermore,  the color distinguishability of bright transitions relies on the hole hybridization energy, the existing Coulomb interactions, and the Zeeman energy shift induced by the magnetic field used for the spin control. The field-dependent photoluminescence spectrum of the QDM is calculated for different transitions contributing in $XX^{-}\rightarrow X^{-}\rightarrow e$ and $XX^{+}\rightarrow X^{+}\rightarrow h$ cascades. These spectra indicate that merely a few spectral components, corresponding to both initial and final favorable spin configurations, are bright. Unfavorable spin configurations refer to those multi-particle states that the Coulomb blockade or Pauli spin blockade prevent them to form. 

The problem of hybrid entanglement between the spin and photons urges coherent control of the electron or hole spin by means of an external magnetic field \cite{Kim2008}. Carrier spin preparation is normally accomplished through exciting the singly-charged QD into the charged exciton ($X^{-}=X+e$ or $X^{+}=X+h$) state and waiting until the spontaneous decay leaves the QD either in $\ket{\uparrow}$ or $\ket{\downarrow}$ states, while the spin state is coherently controlled through a Raman transition driven by a few picosecond pulse at its qubit resonance frequency \cite{Atature2006,Ramsay2008}. Formation of charged biexcitons, $XX^{-}$ and $XX^{+}$, from $\ket{\uparrow}$ or $\ket{\downarrow}$ states is then feasible through a two-photon excitation process \cite{Stufler2006}. We presume that single electron states, $\ket{e_{\uparrow}}$ and $\ket{e_{\downarrow}}$, to be nearly metastable because the $\ket{\uparrow}\leftrightarrows\ket{\downarrow}$
spin-flip transitions are hardly induced by the hyperfine interaction
of the electron spin with the nuclear spin ensemble even at small
magnetic fields ($\gamma_{\ket{\uparrow}\leftrightarrows\ket{\downarrow}}\ll\Gamma$), but are rather regulated by the spin-flip Raman process and the intensity of the stimulating laser \cite{Atature2006,Liu2007}. 

Spectral response of QDM to magnetic field appears in the Zeeman shift of multi-particle states and their spin coupling, thus directly influences the color and linewidth distinguishability of transitions. Along with the Zeeman shift, the exchange interaction causes energy splittings in the spin fine structure of multi-particle states. By performing a detailed configuration interaction (CI) calculation, we however demonstrate that the scale of Zeeman splittings rapidly exceeds the exchange splittings, thus the energy indistinguishability of transitions is primarily violated by Zeeman effect. To obtain a clear sense of the magnetic field intensity required for spin state initialization and manipulation in QDs, we rely on earlier experiments
reporting acceptable fidelities \cite{Press2008,Andlauer2008}. 

A viable solution for erasing these sub-meV energy splittings is to
employ high-efficiency downconversion technique. The method comprises
cross-correlating the emitted photons from QD with a few-picosecond
pulse in a nonlinear medium such as periodically poled lithium niobate
(PPLN) waveguide \cite{Ates2012}. The time resolution of pulses determines
the arrival time of single photons and therefore broadens their linewidths
(lifetime of exciton is commonly beyond 0.5 ns $\gg$ few ps). This
idea could be extended to an arbitrary stream of anti-bunched photons,
including entangled photons. Figure~\ref{fig:QDM shells}(c) schematically shows how the portion of fully entangled photons being delivered by the source can be increased through the linewidth broadening. Twin photons $P_{1}$ and $P_{2}$ have an initial linewidth equivalent to $\Gamma$ and
their color mismatch is $\delta$. Assuming that $\delta$ remains
constant, the overlap region, which implicitly represents the fraction
of quantum correlated photons, grows by enhancing the linewidth $\Gamma\rightarrow\Gamma^{'}$. To quantitatively measure the energy coincidence of photons, we utilize the concept of concurrence developed for entanglement in both regular and time reordering schemes \cite{Wootters1998,Coish2009,Avron2008}.
In the spin fine structure of a typical QD, concurrence is generally
a function of energy mismatches and linewidths of excitonic levels.
We calculate the evolution of concurrence versus the external magnetic field
intensity and photon linewidth. The results reveal that a sufficiently
high concurrence, as compared to the previous experimental results \cite{Hafenbrak2007,Dousse2010}, could be reached by manipulating the linewidths properly. It is worth noting that the QDM with strongly correlated excitons
can be exploited to generate higher order entangled photon states
or hybrid spin-photon states, and the current work serves as a prototypical
study on the application of QDMs in creating tripartite Greenberger-Horne-Zeilinger (GHZ) and W-states \cite{Bouwmeester1999}.

\section{Molecular orbitals in nanowire-double quantum dot}

\begin{figure}
\includegraphics[scale=0.55]{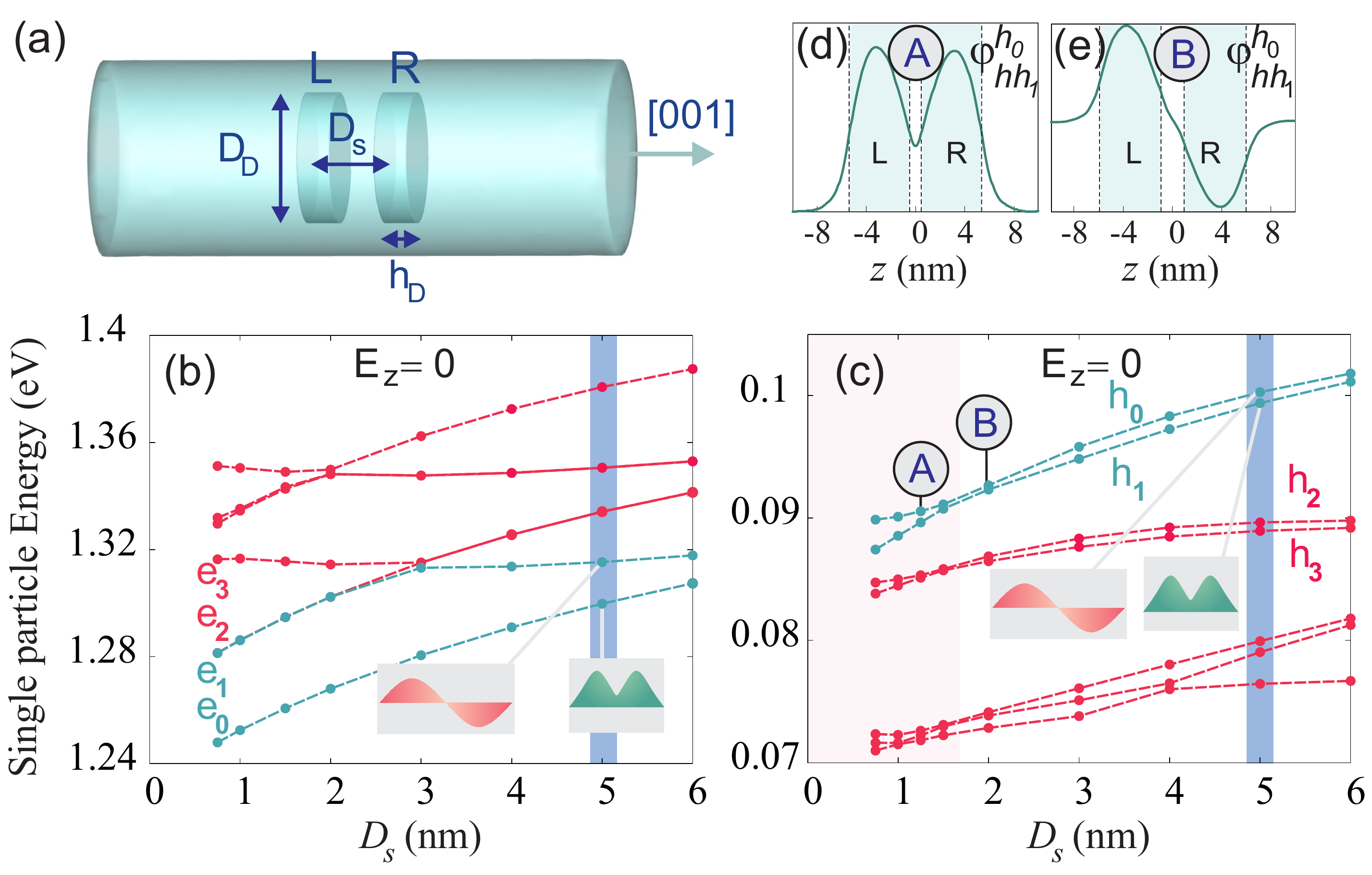}\caption{(a) Schematic of QDM embedded in a [001]-oriented core-shell nanowire. $D_{s}$: the interdot spacing between the left ($L$) and right ($R$) single dots. $h_{D}$: height of the single QD, $D_{D}$: diameter of single QD. Although the structure of QDM is assumed symmetric, the strain field relaxes away from the substrate and becomes axially asymmetric. (b), (c) Electron and hole molecular energies versus the interdot spacing. First and second electron excited states are energetically inseparable for $D_{s}<3$ nm as the whole QDM acts similar to a single dot for electrons at the thin barrier limit. The highlighted region in (c) specifies where the hole ground state is bonding (axially symmetric). Hole tunneling matrix element becomes negative at $D_{s}=1.8$ nm where the hole ground and first excited states anticross. (d), (e) Envelope function of the hole ground state $\varphi_{3/2,3/2}^{h_{0}}$ plotted along the nanowire axis at $D_{s}=1$ and 2 nm showing its bonding and antibonding nature versus $D_{s}$. The highlighted region shows where the hole ground state is symmetric.}

\label{fig:QDM structure_C band_V band}
\end{figure}

The double quantum dot we study consists of two $\textrm{In}{}_{0.5}\textrm{Ga}{}_{0.5}\textrm{As}$
insertions with the ideal $D_{\infty h}$ symmetry embedded inside a
{[}001{]}-oriented GaAs core-shell nanowires; see Figure~\ref{fig:QDM structure_C band_V band}(a). The
QD diameter and its vertical aspect ratio, $a_{h}=h_{D}/D_{D}$, are
chosen 20 nm and 0.25 respectively. This ratio leads to comparable
mutual interactions between electrons and holes, thus the multi-particle
states accumulate less correlation energy \cite{Khoshnegar2012}. The interdot spacing $D_{s}$ determines the wavefunction symmetry of the hole particle ground state $\ket{h_{0}}$ and the sign of its tunneling matrix element $t_{h}$ to hop up to the first excited state $\ket{h_{1}}$. Our numerical calculations
at the single particle level demonstrate that the $\ket{h_{0}}$ symmetry
reversal, as well as the sign inversion of tunneling matrix element \cite{Doty2009} ($t_{h}$ becoming negative), occur at $D_{s}\simeq1.8$ nm. The dominant heavy hole-like part of the ground state envelope function $\varphi_{J=3/2,J_{z}=3/2}^{h_{0}}$
(see Supplemental Information) is illustrated in Figures~\ref{fig:QDM structure_C band_V band}(d-e), where its parity flips moving from $D_{s}=$ 1 to 2 nm.

The molecular ground and first excited states of the hole, $\ket{h_{0}}$
and $\ket{h_{1}}$, are hybridizations of the two dot-localized $s$-shell
states, $\ket{s_{h,R}}$ and $\ket{s_{h,L}}$. These two mixtures
are highly correlated at small double-dot spacing, where $\ket{s_{h,R}}$
and $\ket{s_{h,L}}$ orbitals spread uniformly over the QDM, and become
$\ket{s_{h,R}}$-like or $\ket{s_{h,L}}$-like when the spacing is
sufficiently increased. An analogous situation occurs for the molecular
ground $\ket{e_{0}}$ and first excited $\ket{e_{1}}$ states of the
electron in terms of losing the correlation versus $D_{s}$. Figures~\ref{fig:QDM structure_C band_V band}(b-c) illustrate the evolution of the corresponding electron and hole energies, $E_{e_{0}}$, $E_{e_{1}}$, $E_{h_{0}}$ and $E_{h_{1}}$ against the interdot spacing. A reasonable range of interdot distance can be $4<D_{s}<6$ nm due to the following reasons: 1) hole hybridization energy is suppressed within this range since the tunneling component caused by the spin-orbit interaction weakly exceeds the diagonal tunneling matrix element (the definitions of these tunneling components are explained in Supplemental Information). 2) the electron first excited state (predominantly $s$-like) is noticeably coupled to the second excited state (predominantly $p$-like) given $D_{s}<3$ nm as a result of their trivial energy spacing; this leads to a pronounced correlation between the two molecular states $\ket{e_{1}}$ and $\ket{e_{2}}$ for $D_{s}<3$ nm. 3) $D_{s}$ could be increased limitedly because the hole tunneling matrix element rapidly vanishes. Hereinafter and in order to observe the field-induced spectrum of the molecule at a fixed interdot spacing we set $D_{s}$ equal to 5 nm.

\section{Molecular orbitals versus the axial electric field}

\begin{figure}
\includegraphics[scale=0.8]{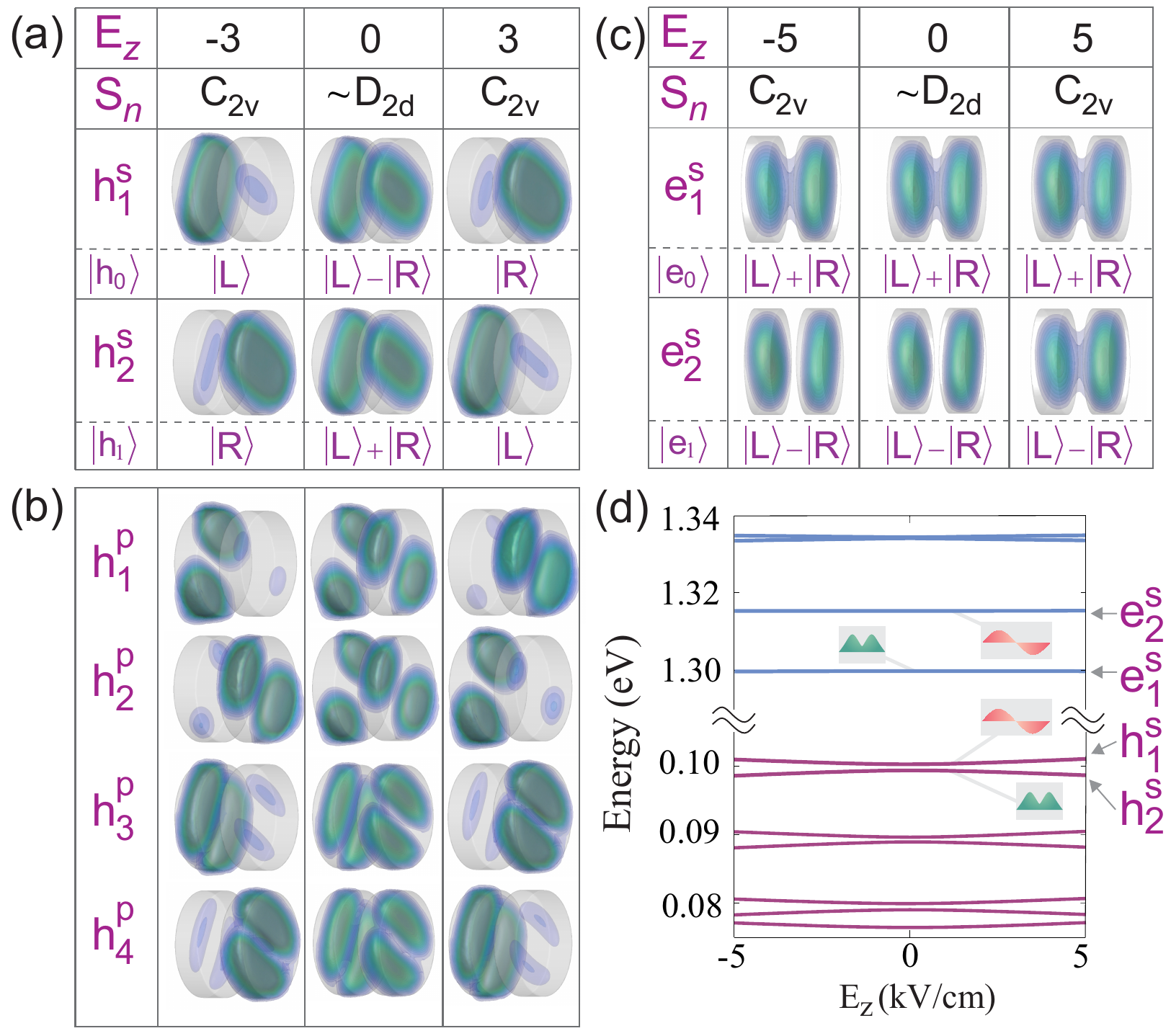}\caption{(a-b) $S$- and $p$-shell hole molecular orbitals versus the electric field indicated in kV/cm. Hole orbitals become localized inside single dots at $E_{z}=\pm 3$ kV/cm, thus their wavefunction could be approximated by the dot-localized states. Applying axial electric field reduces the symmetry character of orbitals from $\sim D_{2d}$ down to $C_{2v}$ (c) Electron molecular orbitals. Owing to the small effective mass of electron, its molecular orbitals reshape reluctantly in response to $E_{z}$. The electron wavefunction remain as the linear mixture of dot-localized states within $-5<E_{z}<5$ kV/cm. The {[}001{]} axis is slightly rotated with respect to the left panel (a and b) in order to illustrate the diffusion of electron orbital inside the GaAs barrier ($t_{e}>t_{h}$). (d) Electron (four states, 2$s$ and 2$p$) and hole (seven states, 2$s$ , 4$p$ and 1$d$) energies as a function of the axial electric field.}

\label{fig:SingleParticleOrbitals}
\end{figure}

In the next step, an axial electric field $E_{z}$ ($z\equiv[001]$)
is applied in order to tailor the absorption energies of the QDM.
Owing to the $D_{\infty h}$ symmetry of single QDs, the orbital densities are rather equivalent once $E_{z}=0$, neglecting trivial
asymmetries caused by the strain-induced fields; see Figure~\ref{fig:SingleParticleOrbitals}(a-c). Both electron and hole molecular orbitals exhibit approximately $D_{2d}$ symmetry, spreading throughout the molecule where each orbital lobe is $C_{2v}$-symmetric. Immediately upon applying the axial electric field, the heavy hole with highest energy $\ket{h_{0}}\equiv\ket{h_{1}^{s}}$ moves toward
lower potential energy, in $\textrm{QD}_{R}$, and effectively evacuates
$\textrm{QD}_{L}$ at $E_{z}\simeq\pm 3\textrm{kV/cm}$ acquiring $C_{2v}$
symmetry; $\ket{h_{0}}=\ket{s_{h,R}}$. The very same situation occurs
for $\ket{h_{1}}\equiv\ket{h_{2}^{s}}$ under negative comparable
electric fields. The Galium intermixing in the QD composition facilitates
both hybridization and interdot diffusion of the electron and hole.
As implied from its dispersion, the electron orbital is stiff against
electric field variations, otherwise it would gain considerable kinetic
energy \cite{Khoshnegar2012}; see Figure~\ref{fig:SingleParticleOrbitals}(c). Thus $\ket{e_{0}}\equiv\ket{e_{1}^{s}}$
and $\ket{e_{1}}\equiv\ket{e_{2}^{s}}$ orbitals sustain their spreadout
over both QDs maintaining the interdot coupling, $\ket{e_{0}}=\ket{s_{e,L}}+\ket{s_{e,R}}$ and $\ket{e_{1}}=\ket{s_{e,L}}-\ket{s_{e,R}}$.

$S$-shell electron and hole energies undergo an anticrossing in proximity
to $E_{z}=0$; see Figure~\ref{fig:SingleParticleOrbitals}(d). In accordance with trivial orbital deformations, the $s$-shell electron energies experience relatively small variations as compared to the $s$-shell hole energies. Along with the relatively large energy spacing between $\ket{e_{0}}$ and $\ket{e_{1}}$, this leads to the predominant hole contribution in any conversion between the excitonic configurations at low electric fields ($|E_{z}|\leq1.5\,\textrm{kV/cm}$). Owing to the strain-induced potentials, electrons and holes do not necessarily localize symmetrically in response to negative and positive electric fields, prohibiting a perfectly symmetric evolution of spectral features versus the axial electric field. 

Since the electrons are weakly localized in our range of electric field, the correct representation is to assign $\ket{P}=\ket{s_{e,L}}+\ket{s_{e,R}}$ and $\ket{N}=\ket{s_{e,L}}-\ket{s_{e,R}}$ to the associated molecular states inward and away from the resonance. In the case of holes, however, we refer to the molecular states as $\ket{R}=\ket{s_{h,R}}$ and $\ket{L}=\ket{s_{h,L}}$ away from the resonance as their orbital could be well approximated by the dot-localized states there. To simplify the nomenclature, we agree to use the $\ket{R}$ and $\ket{L}$ notation for holes even close to the anticrossing, however we are implicitly aware of the fact that $\ket{R}\simeq \ket{s_{h,L}}-\ket{s_{h,R}}$ and $\ket{L}\simeq \ket{s_{h,L}}+\ket{s_{h,R}}$ at this point.

\section{Charged Biexciton Cascades in QDMs: Role of hole tunneling in the energy coincidence of transitions}

\begin{figure}
\includegraphics[scale=0.68]{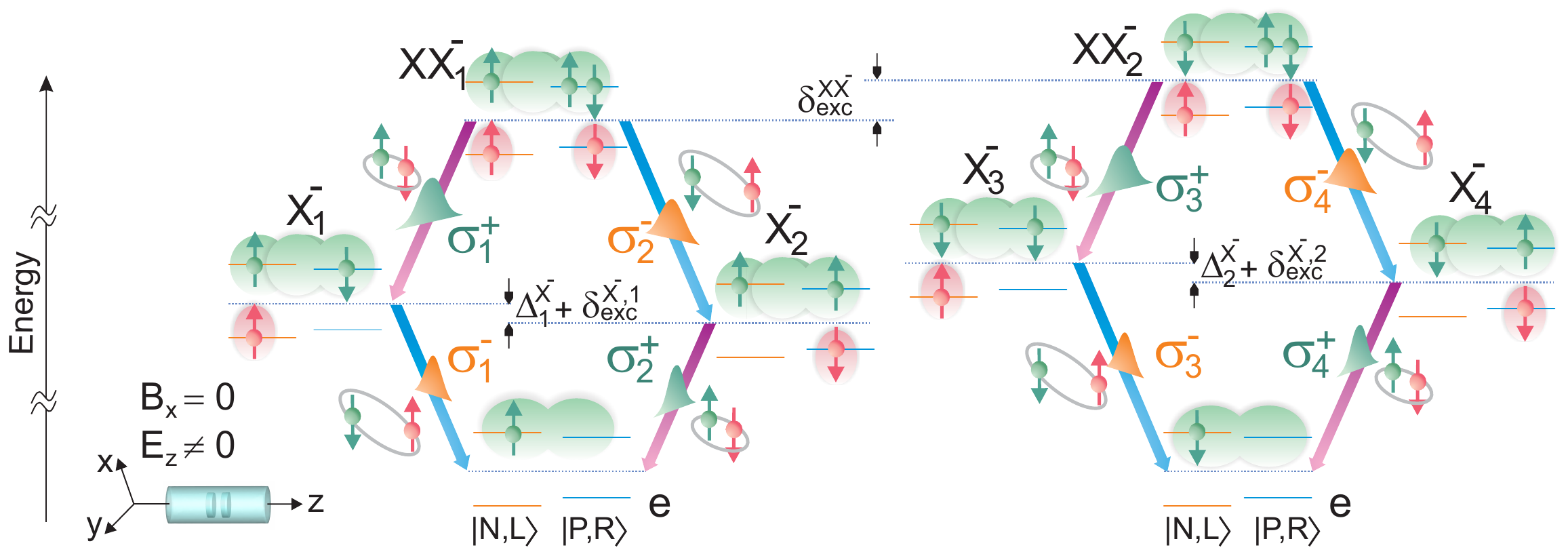}\caption{Cascade decays initiated from $\ket{P_{\uparrow}P_{\downarrow}N_{\uparrow}R_{\Downarrow}L_{\Uparrow}}$
and $\ket{P_{\uparrow}P_{\downarrow}N_{\downarrow}R_{\Downarrow}L_{\Uparrow}}$ negative biexcitons down to intermediate negative trion and electron
states, producing right $\sigma^{+}$ and left $\sigma^{-}$ polarized
photons. The axial electric field is assumed nonzero, $E_{z}\neq 0$, squeezing hole orbitals into dot-localized states. Weak localization of electron is schematically represented by its orbital filling the whole QDM. $\mathbf{B_{\mathit{x}}}=0$, thus spin configurations are still not Zeeman-shifted. Spin-photon pair entanglement could be established between each two paths of the left and right panels (subscripts denote the path numbers). Due to the spin-orbit interaction, degeneracies in the charged configurations are lifted (i.e. nonzero $\delta_{\mathrm{exc}}^{X^{-},1}$, $\delta_{\mathrm{exc}}^{X^{-},2}$ and $\delta_{\mathrm{exc}}^{XX^{-}}$) even at zero magnetic field. $\Delta_{\mathrm{1}}^{X^{-}}+\delta_{\mathrm{exc}}^{X^{-}}$ ($\Delta_{\mathrm{2}}^{X^{-}}+\delta_{\mathrm{exc}}^{X^{-}}$) represents the energy spacing of intermediate trions ruled by the hole tunneling matrix element and Coulomb interactions.}

\label{fig:Cascade_negativeBiexciton}
\end{figure}

Since generating multi-partite correlated states in our scheme relies on the cascade recombination of charged multi-particle complexes, namely $XX^{-}$ and $XX^{+}$, inspecting the contributing transitions of QDM in the presence of hole tunneling sheds light into the essential properties we seek in the energetic of emitted photons. To this end, we consider the exemplary cascades
sketched in Figure~\ref{fig:Cascade_negativeBiexciton} and calculate the transition energies of the diagram. We focus on the cross-entanglement scheme where the hole tunneling energy plays a substantial role. In this scheme, the energy
coincidence between $XX_{k}^{-}\rightarrow X_{i}^{-}$ and $X_{j}^{-}\rightarrow e$
(or $XX_{k}^{+}\rightarrow X_{i}^{+}$ and $X_{j}^{+}\rightarrow h$)
$i\neq j$ transitions must be established; $i$ ($j$) and $k$ sweep over possible spin configurations of $X^{-}$ and $XX^{-}$. In the diagram shown in Figure~\ref{fig:Cascade_negativeBiexciton}, the $XX^{-}$ spin configurations are $\ket{e_{\sigma_{e_{1}}}e_{\sigma_{e_{2}}}e_{\sigma_{e_{3}}}h_{\sigma_{h_{1}}}h_{\sigma_{h_{2}}}}=\ket{P_{\uparrow}P_{\downarrow}N_{\uparrow}R_{\Downarrow}L_{\Uparrow}}$
and $\ket{P_{\uparrow}P_{\downarrow}N_{\downarrow}R_{\Downarrow}L_{\Uparrow}}$,
where one electron molecular state is fully occupied and other states are singly filled up. Upon two inequivalent $\ket{P_{\uparrow}R_{\Downarrow}}$ and $\ket{P_{\downarrow}L_{\Uparrow}}$ recombinations starting from $\ket{P_{\uparrow}P_{\downarrow}N_{\uparrow}R_{\Downarrow}L_{\Uparrow}}$ ($\ket{P_{\uparrow}P_{\downarrow}N_{\downarrow}R_{\Downarrow}L_{\Uparrow}}$), two negative trion states, $\ket{e_{\sigma_{e_{1}}}e_{\sigma_{e_{2}}}h_{\sigma_{h_{1}}}}=\ket{P_{\downarrow}N_{\uparrow}L_{\Uparrow}}$ ($\ket{P_{\downarrow}N_{\downarrow}L_{\Uparrow}}$) and $\ket{P_{\uparrow}N_{\uparrow}R_{\Downarrow}}$ ($\ket{P_{\uparrow}N_{\downarrow}R_{\Downarrow}}$), emerge whose energy spacing, labeled by $\Delta_{1}^{X^{-}}+\delta_{\mathrm{exc}}^{X^{-},1}$ ($\Delta_{2}^{X^{-}}+\delta_{\mathrm{exc}}^{X^{-},2}$) in Figure~\ref{fig:Cascade_negativeBiexciton}, lies in the hole tunneling matrix element along with the direct and exchange Coulomb interactions. The absorption energy detuning between $\sigma_{1}^{+}$-photon in path $\mathcal{P_{\mathrm{1}}}$ and $\sigma_{4}^{+}$-photon in path $\mathcal{P_{\mathrm{4}}}$ under zero magnetic field, $\delta^{\sigma_{1,4}^{+}}$ ($\approx\delta^{\sigma_{4,1}^{-}}$, neglecting the difference between exchange interactions), is given by

\begin{equation}
\delta^{\sigma_{1,4}^{+}}=2t_{h}+J_{hh}^{LR}+J_{ee}^{NN}-J_{eh}^{NR}-J_{eh}^{NL}+\Delta_{\mathrm{corr}}^{XX_{1}^{-}}-\Delta_{\mathrm{corr}}^{X_{1}^{-}}-\Delta_{\mathrm{corr}}^{X_{4}^{-}}+\Delta_{\mathrm{exc}}^{XX_{1}^{-}}-\Delta_{\mathrm{exc}}^{X_{1}^{-}}-\Delta_{\mathrm{exc}}^{X_{4}^{-}},\label{eq:4-1}
\end{equation}
where $t_{h}=(E_{h}^{L}-E_{h}^{R})/2$ represents the hole tunneling matrix element,  $J_{ab}^{AB}$ stands for the mutual direct Coulomb interaction between particles $a$ and $b$ in the molecular states $A$ and $B$,
and $\Delta_{\mathrm{corr}}^{\alpha}$ ($\Delta_{\mathrm{exc}}^{\alpha}$)
denote the correlation (exchange) energy stored in the complex $\alpha$.
Superscript $AB$ denotes the potential of a single particle resting
in molecular state $A$ being felt by another single particle localized in state $B$. In order to write down above equation, the total energies of initial
and final complexes in each transition are derived and subtracted.
Eq.~\ref{eq:4-1} holds once the two paths (here $\mathcal{P_{\mathrm{1}}}$
and $\mathcal{P_{\mathrm{4}}}$) do not share identical initial (negative
biexciton) and ground (electron) states.

As implied by Eq.~\ref{eq:4-1}, $\delta^{\sigma_{1,4}^{+}}$ can be suppressed
upon comparable attractive ($J_{eh}^{AB}$) and repulsive ($J_{hh}^{AB}$
or $J_{ee}^{AB}$) Coulomb interactions assuming $t_{h}$, $\Delta^{\alpha}_{\mathrm{corr}}$, $\Delta^{\alpha}_{\mathrm{exc}}\ll J^{AB}_{ab}$ \cite{Schliwa2009}. Fulfilling the condition $J_{hh}^{LR}+J_{ee}^{NN}=J_{eh}^{NR}+J_{eh}^{NL}$ then requires a relative similarity between the electron molecular states as well as left- and right-localized hole states. The necessity of this resemblance further unfolds when considering a single QD: $XX^{-}$ forms once a biexciton fills up the $s$-shell and a single electron resides the $p$-shell.
Since transitions between the $s$ and $p$ shells are ideally forbidden
$M_{sp}\ll M_{ss},\, M_{pp}$ ($M_{AB}$ stands for the oscillator strength of transition between the states $A$ and $B$),
either (a) one recombination within each path is optically unfavorable
or (b) the final metastable state varies between $s$ and $p$ shell,
hindering the energy coincidence. In the QDM proposed here, this coincidence
is likely to occur merely when the two decaying paths share the same
ground state, $\ket{N_{\uparrow,\downarrow}}$ or $\ket{P_{\uparrow,\downarrow}}$, as the electron tunneling matrix element exceeds few meVs.

All charged configurations in Figure~\ref{fig:Cascade_negativeBiexciton} further split after introducing a magnetic field in the Voigt geometry (transverse to the primary quantization axis, $\mathbf{B}_{x}\neq0$). The induced
energy splitting $\Delta_{\mathbf{B}}^{\alpha_{i},\alpha_{j}}$
between charged configurations $\alpha_{i}$ and $\alpha_{j}$ then
relies on their Zeeman mixing renormalized by the few-body correlations.
The order of this splitting under adequately intense magnetic field
($>200\,\mu\mathrm{eV}$) is large enough compared to the dephasing
linewidth of $\sigma^{+}$ (or $\sigma^{-}$) transition to impair
the certainty in the phase of spin-photon pair wavefunction. Applying
weak magnetic fields, on the other hand, could not maintain enough
fidelity of the few-photon-based spin readout \cite{Press2008}. Notice
that in addition to the recombinations shown in Figure~\ref{fig:Cascade_negativeBiexciton}, dark excitons ($e_{\uparrow}h_{\Uparrow}$ or $e_{\downarrow}h_{\Downarrow}$) become optically active in the Voigt geometry and give rise to extra cascade combinations; this is further detailed in Sec. VI. Above criteria could be safely applied to the other existing $XX^{-}\rightarrow X^{-}\rightarrow e$ cascades and also to $XX^{+}\rightarrow X^{+}\rightarrow h$ cascades. 

Accounting only for the bright transitions, the spin-photon
pair entangled states from paths $\mathcal{P_{\mathrm{1,4}}}$ and
$\mathcal{P_{\mathrm{2,3}}}$ are 

\begin{equation}
1/\sqrt{2}(\ket{\sigma^{+}(\omega_{1})\sigma^{+}(\omega_{4})}\otimes\ket{\downarrow}+\ket{\sigma^{-}(\omega_{4})\sigma^{-}(\omega_{1})}\otimes\ket{\uparrow})
\end{equation}

and 

\begin{equation}
1/\sqrt{2}(\ket{\sigma^{-}(\omega_{2})\sigma^{-}(\omega_{3})}\otimes\ket{\downarrow}+\ket{\sigma^{+}(\omega_{3})\sigma^{+}(\omega_{2})}\otimes\ket{\uparrow}),
\end{equation}
respectively, where $\hbar\omega_{1}^{\pm}=\hbar\omega_{4}^{\pm}\pm\delta^{\sigma_{1,4}^{+}}+\mathcal{O}{}_{1}(\delta_{\mathrm{exc}},\Delta_{\mathbf{B}})=\hbar\omega_{3}^{\pm}+\mathcal{O}{}_{2}(\delta_{\mathrm{exc}},\Delta_{\mathbf{B}})=\hbar\omega_{2}^{\pm}\pm\delta^{\sigma_{1,4}^{+}}+\mathcal{O}{}_{3}(\delta_{\mathrm{exc}},\Delta_{\mathbf{B}})$, and $\mathcal{O}(\delta_{\mathrm{exc}},\Delta_{\mathbf{B}})$ represents
the energy correction due to the exchange and Zeeman splittings. As offered by Figure~\ref{fig:SingleParticleOrbitals}(d), the hole anticrossing and thus the negative trion splittings, $\Delta_{1}^{X^{-}}$ and $\Delta_{2}^{X^{-}}$, are in the order of 0.9 meV ($t_{h}<0.45$ meV) at $D_{s}$= 5 nm. This small tunneling energy allows for the
hole to readily commute between the dots and render higher energy
resolutions in the detuning $\delta^{\sigma_{1,4}^{+}}$ adjustment.

\section{Photoluminescence spectra of charged complexes under electric field}
\label{sec:VI}

In this section, we present the results of CI calculations incorporating
the Coulomb direct and exchange interactions in a universal few-body
hamiltonian constructed from the single particle orbitals (see Supplemental
Information). Table~\ref{Table:excitonic_classes} shows different classes of excitonic complexes studied here. Consider the case that electron molecular orbitals were strongly localized in different dots: $\ket{P}=\ket{s_{e,R}}$ and $\ket{N}=\ket{s_{e,L}}$. The $s$-shell transitions then could be categorized into direct and indirect as shown in Table~\ref{Table:excitonic_classes}. The indirect transitions are switched off in such a QDM hosting weakly correlated excitons. In a relatively symmetric QDM with weakly localized electrons, however, they have the chance to recombine with either left- or right-localized holes depending on the total energy of the final complex or particle. This will be detailed in the following.

\begin{table}
\caption{Labels and number of $s$-shell configurations of multi-particle complexes studied here. All four complexes $X^{-}$, $X^{+}$, $XX^{-}$ and
$XX^{+}$ have six spin-excluded components free of exchange interactions.
Two last columns show the number of direct, $e_{\uparrow}^{P|N}h_{\Downarrow}^{R|L}$ or $e_{\downarrow}^{P|N}h_{\Uparrow}^{R|L}$, and indirect, $e_{\uparrow}^{N|P}h_{\Downarrow}^{R|L}$ or $e_{\downarrow}^{N|P}h_{\Uparrow}^{R|L}$, bright excitons existing in each complex. }
\begin{tabular}{llclcccc}
\hline 
\multicolumn{1}{l}{} & \multicolumn{1}{c}{} & \multicolumn{2}{c}{} & \multicolumn{4}{c}{$s$-shell bright channels}\tabularnewline
 &  &  &  & \multicolumn{4}{c}{(Spin included)}\tabularnewline
\cline{4-8} 
\multirow{2}{*}{Complex} & \multirow{2}{*}{Label} & $s$-shell configs. &  & Total & Direct &  & Indirect\tabularnewline
 &  & (No spin) &  &  &  &  & \tabularnewline
N-trion ($X^{-}$) & $eeh$ & 6 & $X^{-}\rightarrow e$ & 24 & 12 &  & 12\tabularnewline
P-trion ($X^{+}$) & $hhe$ & 6 & $X^{+}\rightarrow h$ & 24 & 12 &  & 12\tabularnewline
N-biexciton ($XX^{-}$) & $eeehh$ & 6 & $XX^{-}\rightarrow X^{-}$ & 72 & 36 &  & 36\tabularnewline
P-biexciton ($XX^{+}$) & $hhhee$ & 6 & $XX^{+}\rightarrow X^{+}$ & 72 & 36 &  & 36\tabularnewline
\hline 
\end{tabular}

\label{Table:excitonic_classes}
\end{table}

\subsection{Negatively charged exciton, $X^{-}\rightarrow e$ transition}

Neglecting the spin degree of freedom, 12 transitions could occur between the negative exciton and the single electron levels ($6X^{-}\times2e$). Among these, a part of tunneling-assisted transitions are ruled out depending on how the single particles redistribute their orbitals according to the Coulomb interactions. The formation mechanism of $X^{-}$ consists of initializing one electron spin in the ground state followed by a photoexcitation creating an electron-hole pair. The favorable configuration for the two electrons is to occupy different molecular states $\ket{P}$ and $\ket{N}$ as a consequence of repulsive interaction. $\ket{P}$ and $\ket{N}$, however, remain as
a strong superposition of $\ket{s_{e,L}}$ and $\ket{s_{e,R}}$ in
our range of electric field. The hole particle may exist either in
the left or in the right QD, whether it is energetically favorable
or not. For the sake of simplicity, hereafter, we represent $\ket{s_{e,L}}$ and $\ket{s_{e,R}}$ by $\ket{L}$ and  $\ket{R}$, respectively.

\begin{figure}
\includegraphics[scale=0.6]{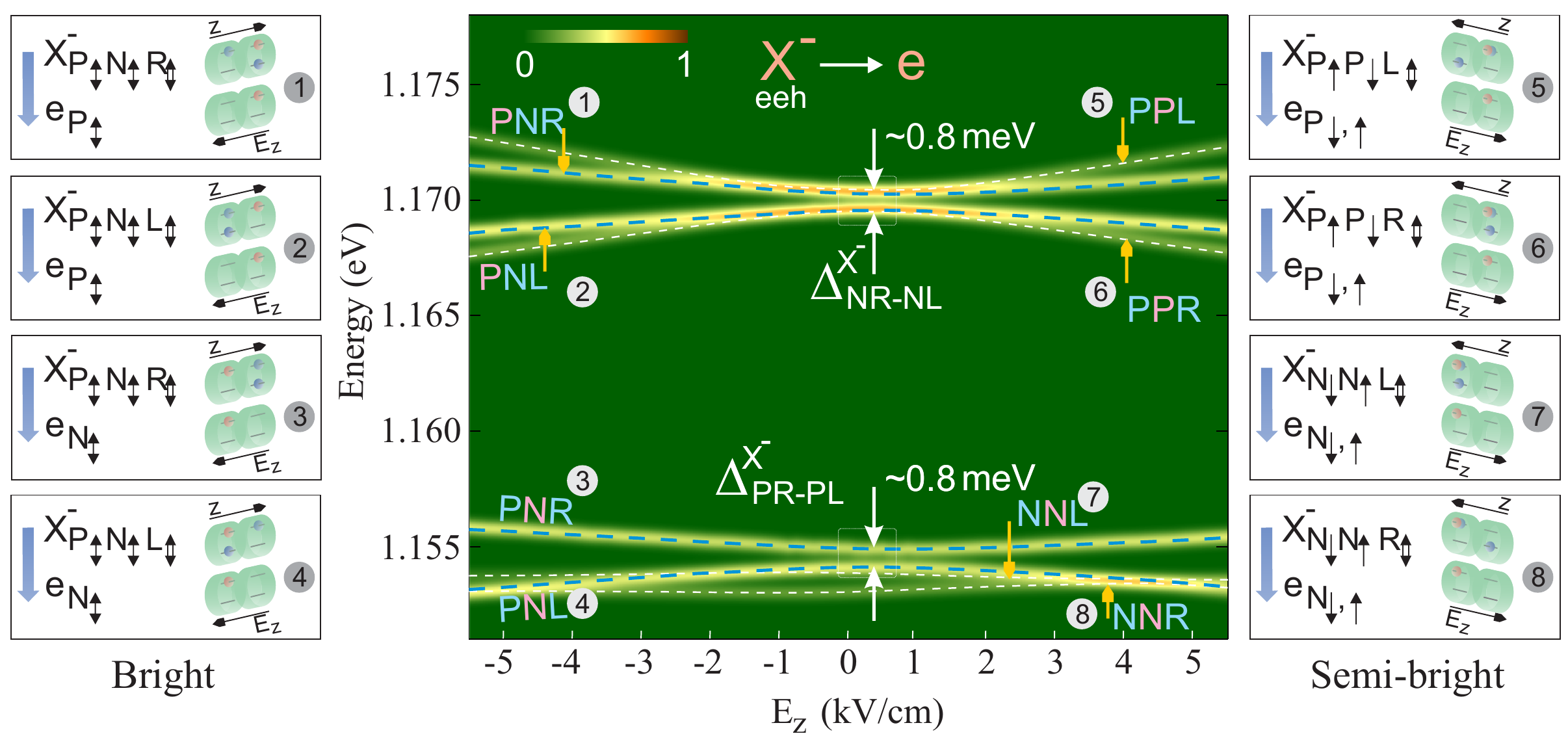}\caption{Left and right panels: electron and hole configurations in the bright and semi-bright spectral components of negative trion $X^{-}$. Center: photoluminescence intensity of $X^{-}$ bright and semi-bright components plotted against the axial electric field and $X^{-}\rightarrow e$ transition energy. Recombining particles are coloured blue. The averaged energies associated with different components are highlighted by dashed lines: blue, favorable bright configurations; white, semi-bright configurations. Squares show the region of trion-bound hole anticrossing: $\Delta_{NR-NL}^{X^{-}}$ and $\Delta_{PR-PL}^{X^{-}}\approx0.8\,\mathrm{meV}$ comparable to the magnitude of hole anticrossing gap; the subscript shows the recombining excitons in each anticrossing.}

\label{fig:XN_to_e}
\end{figure}

Figure~\ref{fig:XN_to_e} illustrates the photoluminescence intensity of $X_{e}^{-}$($\equiv X^{-}\rightarrow e$) transitions versus the axial electric field and the associated energies. The transition from the initial configuration $i$ to the final configuration $f$ is represented by $i_{f}$ and all spectral components are plotted with reference energy located at the QDM center. A relatively large energy spacing between $\ket{P}$ and $\ket{L}$
translates to two families of transitions appearing in the spectrum.
Among all the features, four components around 1.155 eV and 1.17 eV
exhibit bright photoluminescence intensity. a) Bright transitions:
$X^{-}\{eeh\}_{e\{e\}}\equiv$ $PNR_{N}$, $PNL_{N}$,
$PNR_{P}$ and $PNL_{P}$ for $E_{z}<0.5\,\mathrm{kV/cm}$.
Note that in this representation, particles flip their localization
($R\rightleftharpoons L$) once the electric field is reversed, except
those particles undergoing an anticrossing. We observe no pronounced
distinction between the $M_{NR}$ ($M_{PL}$) and $M_{NL}$ ($M_{PR}$) oscillator strengths. This primarily relates to the conformity
of electron configurations in the two initial states of these transitions bearing a hole anticrossing, and also to the extension of electron orbitals spreading over the entire molecule: $PNR_{N}$ and $PNL_{N}$ transitions are coupled sharing the same final state ($\ket{N}$ for $E_{z}<0.5\,\mathrm{kV/cm}$) and $\braket{P|R}\approx\braket{P|L}$ remains unaffected upon heavy hole tunneling.

The anticrossing energy $\Delta_{PR-PL}^{X^{-}}\approx0.8\,\mathrm{meV}$ at $E_{z}=0.5\,\mathrm{kV/cm}$ is comparable to twice the heavy hole tunneling matrix element $\sim0.86\,\mathrm{meV}$ (see Supplemental Information) but has been slightly renormalized by the Coulomb correlations accumulated in $X^{-}$. b) Semi-bright transitions, including $NNL_{N}$, $NNR_{N}$, $PPL_{P}$ and $PPR_{P}$ ($E_{z}>0.5\,\mathrm{kV/cm})$, have two electrons residing the same molecular state in their initial trion. The photoexcited $X^{-}$ state is then unfavorable to form, thus the oscillator strength drops significantly regardless of whether recombining particles resting in the same dot or not. c) Dark transitions, such as $PPR_{N}$, exist where one electron post-tunneling is involved after recombination (not shown here). We attribute the asymmetry evident in the trion features mostly to the modest asymmetry of the electron and hole orbitals in response to oppositely-oriented axial fields.

\subsection{Negatively charged biexciton, $XX^{-}\rightarrow X^{-}$ transition}

\begin{figure}
\center
\includegraphics[scale=0.6]{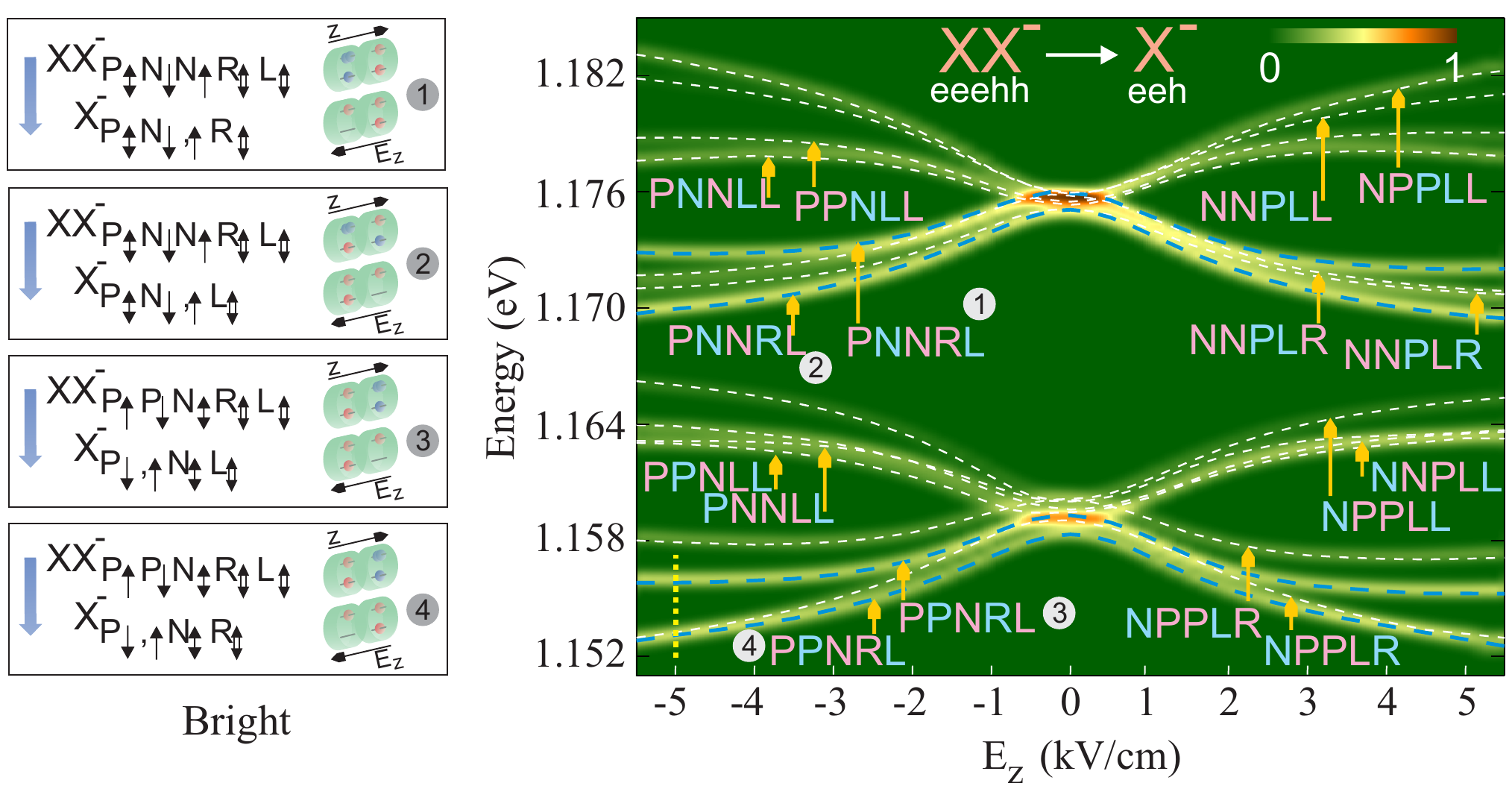}\caption{Left: electron and hole configurations in the bright spectral components of negatively charged biexciton $XX^{-}$. Right: photoluminescence intensity of  $XX^{-}$ versus the axial electric field. Blue (white) dashed lines identify bright (semi-bright) transitions. There exist four bright components, in which both initial ($XX^{-}$) and final ($X^{-}$) configurations are favorable to form: $PNNRL_{PNR}$, $PNNRL_{PNL}$, $PPNRL_{PNL}$ and $PPNRL_{PNR}$. By comparing to Figure \ref{fig:XN_to_e}, we realize that $\delta^{\sigma_{1,4}^{+}}$ ($\delta^{\sigma_{1,4}^{-}}$) is minimized at $E_{z}=-5\,\mathrm{kV/cm}$ (marked by yellow dotted line) where $PPNRL_{PNR}$ ($PPNRL_{PNL}$) and $PNL_{N}$ ($PNR_{N}$) coincide.}

\label{fig:XXN_to_XN}
\end{figure}

In contrast to $X^{-}$, favorable $XX^{-}$ configurations accommodate
two holes separated in different dots while one electron molecular state
is fully and the other one is singly populated. Figure~\ref{fig:XXN_to_XN} shows the two transition families experiencing
the hole anticrossings around 1.158 eV and 1.175 eV. Analogous to
$X^{-}$, the bright components comprise the following recombinations: $XX^{-}\{eeehh\}_{X^{-}\{eeh\}}\equiv$ $PNNRL_{PNL}$, $PNNRL_{PNR}$, $PPNRL_{PNR}$ and
$PPNRL_{PNL}$. A part of the semi-bright components originate from
those configurations having two holes in the same dot, that is $XX^{-}\{eeehh\}_{X^{-}\{eeh\}}\equiv$ $eeeRR_{eeR}$ or $eeeLL_{eeL}$, where the unlabeled
electron could choose both $\ket{P}$ and $\ket{N}$ molecular states. The rest form once the fully-populated electron molecular orbital of the trion state reshapes along the direction of electric field, rather than in the opposite direction. Some
features, including bright components, are subjected to sizeable energy
shifts up to 8 meV as a function of the electric field. All $XX^{-}$
configurations comprising separated holes rise in energy once the
electric field is suppressed since $J_{hh}^{LR}$ increases. This
increase returns to the significant enhancement of the $\braket{h_{0}|h_{1}}$ spatial overlap despite the quick decrease in the $\ket{R}$ and $\ket{L}$ orbital densities. In those components where the two holes populate the same QD, $J_{hh}^{LL,RR}$ drops at smaller electric fields because both $\braket{h_{0;\Uparrow}|h_{0;\Downarrow}}$ ($\braket{h_{1;\Uparrow}|h_{1;\Downarrow}}$) and the orbital densities diminish.

We focus on the two lower-energy bright $XX^{-}\rightarrow X^{-}$ transitions previously illustrated in Figure~\ref{fig:Cascade_negativeBiexciton}, $PPNRL_{PNR}$ and $PPNRL_{PNL}$. Assuming exactly identical initial configurations (spin included), the two transitions are tunnel coupled through their final trion states and separated by $\sim\Delta_{h}^{X^{-}}=\Delta_{NR-NL}^{X^{-}}$ splitting at $E_{z}=0.5\,\mathrm{kV/cm}$. The highlighted energy
lines in Figure~\ref{fig:XXN_to_XN}, however, merely indicate the average energy of
each component regardless of its spin fine structure. 
Comparing Figures~\ref{fig:XN_to_e} and \ref{fig:XXN_to_XN}, we notice that $PPNRL_{PNR}$ ($PPNRL_{PNL}$) and $PNL_{N}$ ($PNR_{N}$) transitions almost energetically coincide
at $E_{z}=-5\,\mathrm{kV/cm}$. Neglecting the correlation energies
in Eq. \ref{eq:4-1}, $J_{hh}^{LR}$ is subjected to $\sim5\,\mathrm{meV}$
(50.25 to 46.25 meV) reduction in magnitude due to the $h_{0}$-$h_{1}$
spatial splitting caused by a negative electric field equivalent to
$5\,\mathrm{kV/cm}$. $J_{ee}^{NN}$ remains almost unchanged (48.5
meV), while $J_{eh}^{NL}$ and $J_{eh}^{NR}$ each undergo $-0.4$
meV and $1.6$ meV shift. The size of QDM and the interdot spacing
allow for less confinement along the angular momentum quantization
axis $z$ as compared to the previously reported single or stacked
QDs \cite{Bester2004,Ramirez2010}. The moderate axial confinement
then leads to the smooth variation of $h$-$h$ interaction while
$e$-$h$ and $e$-$e$ interactions stay roughly fixed. Although
$\ket{h_{0}}$ and $\ket{h_{1}}$ molecular orbitals are spatially separated at such
fields, their excitons are correlated via the bound electrons. 

\subsection{Positively charged complexes, $X^{+}\rightarrow h$ and $XX^{+}\rightarrow X^{+}$ transition}

\begin{figure}
\center
\includegraphics[scale=0.6]{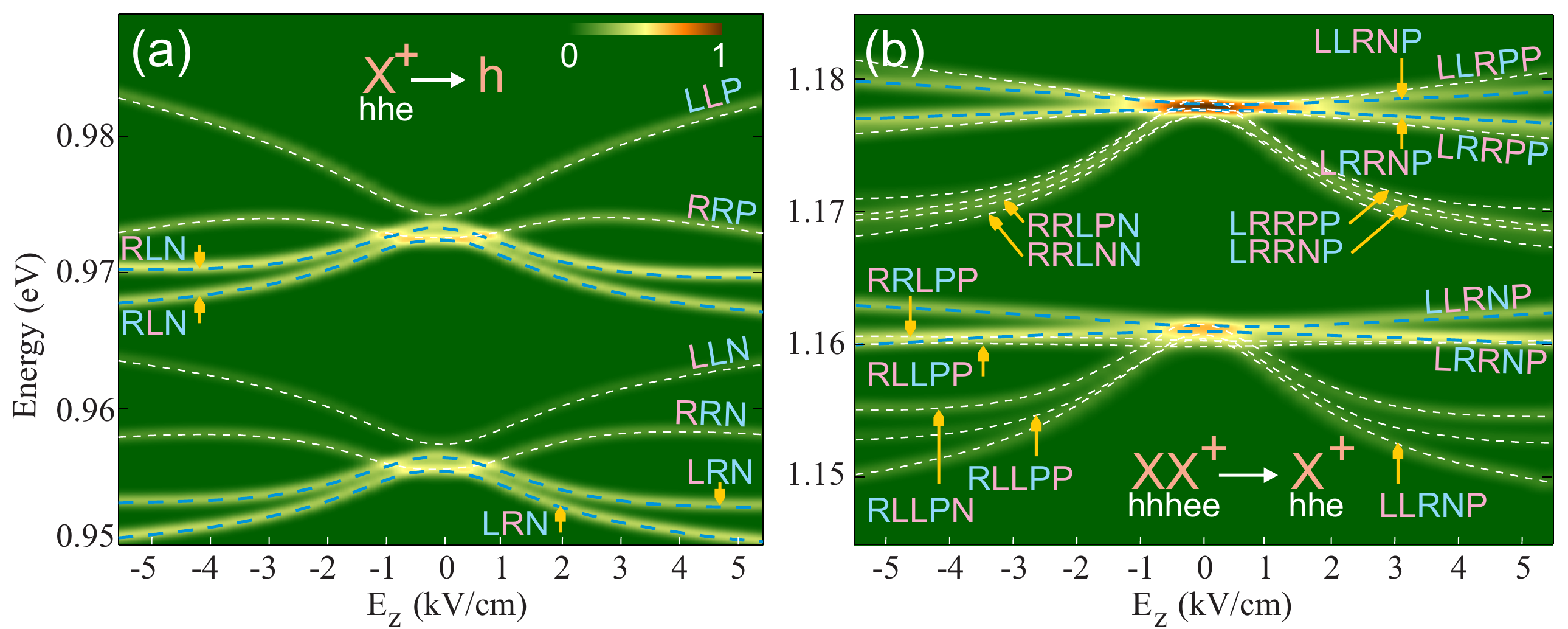}\caption{Photoluminescence intensity depicted for (a) positive trion $X^{+}$ and (b) positive biexciton $XX^{+}$ against the axial electric field and transition energies. Bright components in $X^{+}$ spectrum form with spatially separated holes. The $XX^{+}\rightarrow X^{+}$ spectrum is, by average, 20 meV blueshifted with respect to the $X^{+}\rightarrow h$ bright transitions due to the electron hybridization energy. Bright components of $XX^{+}$ include initial configurations with separated electrons and final configurations $X^{+}$ with separated holes: $LLRNP_{LRN}$, $LRRNP_{LRN}$, $LLRNP_{LRP}$ and $LRRNP_{LRP}$. Semi-bright transitions in $XX^{+}\rightarrow X^{+}$ comprise either two neighboring electrons in the initial state or two neighboring holes in the final state. The energy splitting between the $RLN_{L}$  and $RLN_{R}$ ($LRN_{L}$ and $LRN_{R}$) components at small fields primarily correspond to hole anticrossing in their final states.}

\label{fig:XXP_to_XP_to_h}
\end{figure}

Analogous to the negative trion, formation of a positive trion consists of spin initialization and photoexcitation, but the hole spin is coherently controlled here. Favorable
arrangements include the electron neighboring the initial hole whereas
the subsequent hole moving to the other dot. This is implied by Figure~\ref{fig:XXP_to_XP_to_h}(a) where the bright features correspond to separated holes: $X^{+}\{hhe\}_{h\{h\}}\equiv$
$LRN_{L}$, $LRN_{R}$ ($E_{z}>0$), $RLN_{R}$ and $RLN_{L}$ ($E_{z}<0$).
Regardless of the number of transitions, small energy corrections,
and the spectral redshift, we seek band bendings akin the $XX^{-}\rightarrow X^{-}$
transitions since the energy evolution of each component versus the
electric field is primarily attributed to the changes in $J_{hh}^{LR}$
($J_{hh}^{LL}$ or $J_{hh}^{RR}$ in semi-bright features). In contrast
to $XX^{-}$ and $X^{+}$, $XX^{+}$ contains three interacting hole
particles, complicating the photoluminescence spectrum. The highly favorable configurations
develop with segregated electrons, however, the $e$-$e$ repulsion
does not severely influence the formation of pair electrons in our
range of axial field, because the energy difference between $P_{\uparrow}$-$P_{\downarrow}$
(or $N_{\uparrow}$-$N_{\downarrow}$) and $P_{\updownarrow}$-$N_{\updownarrow}$
repulsions is insignificant. Transitions in Figure~\ref{fig:XXP_to_XP_to_h}(b) can be divided
into two classes based on their final trion state: a) components in
which the hole particles in the final state live in the same dot. These
transitions are more responsive toward $h$-$h$ repulsion compared
to b) components with segregated holes in their final trion, thus bearing
higher energy shift at larger fields. No features from $XX^{+}\rightarrow X^{+}$
and $X^{+}\rightarrow h$ spectra coincide due to the relatively large
hybridization energy of the extra electron living in $XX^{+}$. This
signifies that multipartite energy matching via both negatively and
positively charged complexes is, in general, feasible in QDMs where
electron and hole tunneling matrix elements are tantamount \cite{Peng2010}.

In the final part of this section, let us concisely address the case
of the {[}001{]} nanowire replaced with its {[}111{]} counterpart.
The strain-driven asymmetry of single particle states seen above,
and also the asymmetry of transition energies versus the axial electric
field are more pronounced in {[}111{]}-oriented double QDs as a result
of the built-in piezoelectric field \cite{Niquet2008}. This internal
axial piezoelectric field separates $\ket{h_{0}}$ and $\ket{h_{1}}$
molecular orbitals even at $E_{z}\approx0$, further splitting the locations
of electron and hole anticrossings at low electric fields. The advantage
of {[}111{]}-oriented quantization, however, is that orbitals save
their $C_{3v}$ symmetry irrespective of the electric field amplitude.
As mentioned above, the symmetry conversion from $D_{2d}$ down to
$C_{2v}$ upon applying the axial electric field on {[}001{]}-QDMs
lifts the bright exciton degeneracies. In contrast, the elevated
$C_{3v}$ symmetry predicted in {[}111{]}-QDMs is immune to the symmetry-breaking
Stark effect. 

\section{Analytical hamiltonian: effect of multi-particle interactions on the spin fine structures}

In this section, we develop a simplified version of the $XX^-$ and $X^{-}$
hamiltonians to comprehend the effect of Coulomb interactions on the spin-conserved tunnelings. Providing an analytical model aids us to visualize the behavior of $XX^-$ and $X^{-}$ spin fine structure versus electric field, without getting involved with the details of CI method. We confine our study
to the favorable configurations of $XX^-$ and $X^{-}$ shown in Figure~\ref{fig:Cascade_negativeBiexciton}. For simplicity, we consider the $e$-$h$ exchange interactions of single particles residing different molecular states equal, $\delta_{\mathrm{exc}}^{NL}=\delta_{\mathrm{exc}}^{PR}=\delta_{\mathrm{exc}}^{NR}=\delta_{\mathrm{exc}}^{PL}=\delta_{\mathrm{exc}}^{eh}$. This assumption is valid within our range of electric fields where
electron and hole molecular orbitals consistently maintain their
overlap integral and the associated mean field potentials rather unchanged. The $4\times4$ few-body hamiltonian
describing the non-degenerate bright configurations of the negative
trion $X_{eeh=PNh}$ could be expanded on basis states $N_{\uparrow}P_{\downarrow}L_{\Uparrow}$, $N_{\uparrow}P_{\downarrow}R_{\Uparrow}$, $N_{\downarrow}P_{\downarrow}L_{\Uparrow}$ and $N_{\downarrow}P_{\downarrow}R_{\Uparrow}$

\begin{equation}
\mathcal{H}_{\mathrm{\mathit{X^{-}}}}^{\mathrm{B}}=\left(\begin{array}{cccc}
\zeta_{L}^{X^{-}} & -t_{h} & 0 & 0\\
-t_{h} & \zeta_{L}^{X^{-}}+\Delta J^{X^{-}}-\zeta_{E_{z}} & 0 & 0\\
0 & 0 & \zeta_{L}^{X^{-}}+2\delta_{\mathrm{exc}} & -t_{h}\\
0 & 0 & -t_{h} & \zeta_{L}^{X^{-}}+\Delta J^{X^{-}}-\zeta_{E_{z}}+2\delta_{\mathrm{exc}}
\end{array}\right).\label{eq:negativeTrionHamiltonian}
\end{equation}

Here, $\zeta_{L}^{X^{-}}$ is the total energy of $N_{\uparrow}P_{\downarrow}L_{\Uparrow}$, $\zeta_{E_{z}}$ signifies the energy shift caused by the electric field when the reference energy is located at center of the left QD, $\zeta_{E_{z}}=\overline{e}E_{z}(D_{s}+h_{D})$, $\overline{e}$ is the unit charge, and $\Delta J^{X^{-}}=J_{eh}^{PL}+J_{eh}^{NL}-J_{eh}^{NR}-J_{eh}^{PR}$. The slight difference between $s$-shell molecular orbitals leads to nonzero $\Delta J^{X^{-}}$. The mean value of the $e$-$h$ exchange energy $\delta^{eh}_{\mathrm{exc}}$ is calculated by averaging over $\bra{\psi_{\uparrow}^{e}\psi_{\Downarrow}^{h}}\mathcal{C}\ket{\psi_{\Downarrow}^{h}\psi_{\uparrow}^{e}}-\bra{\psi_{\uparrow}^{e}\psi_{\Uparrow}^{h}}\mathcal{C}\ket{\psi_{\Uparrow}^{h}\psi_{\uparrow}^{e}}$ versus $E_{z}$, where $\mathcal{C}$ is the Coulomb interaction operator.

\begin{figure}
\includegraphics[scale=0.5]{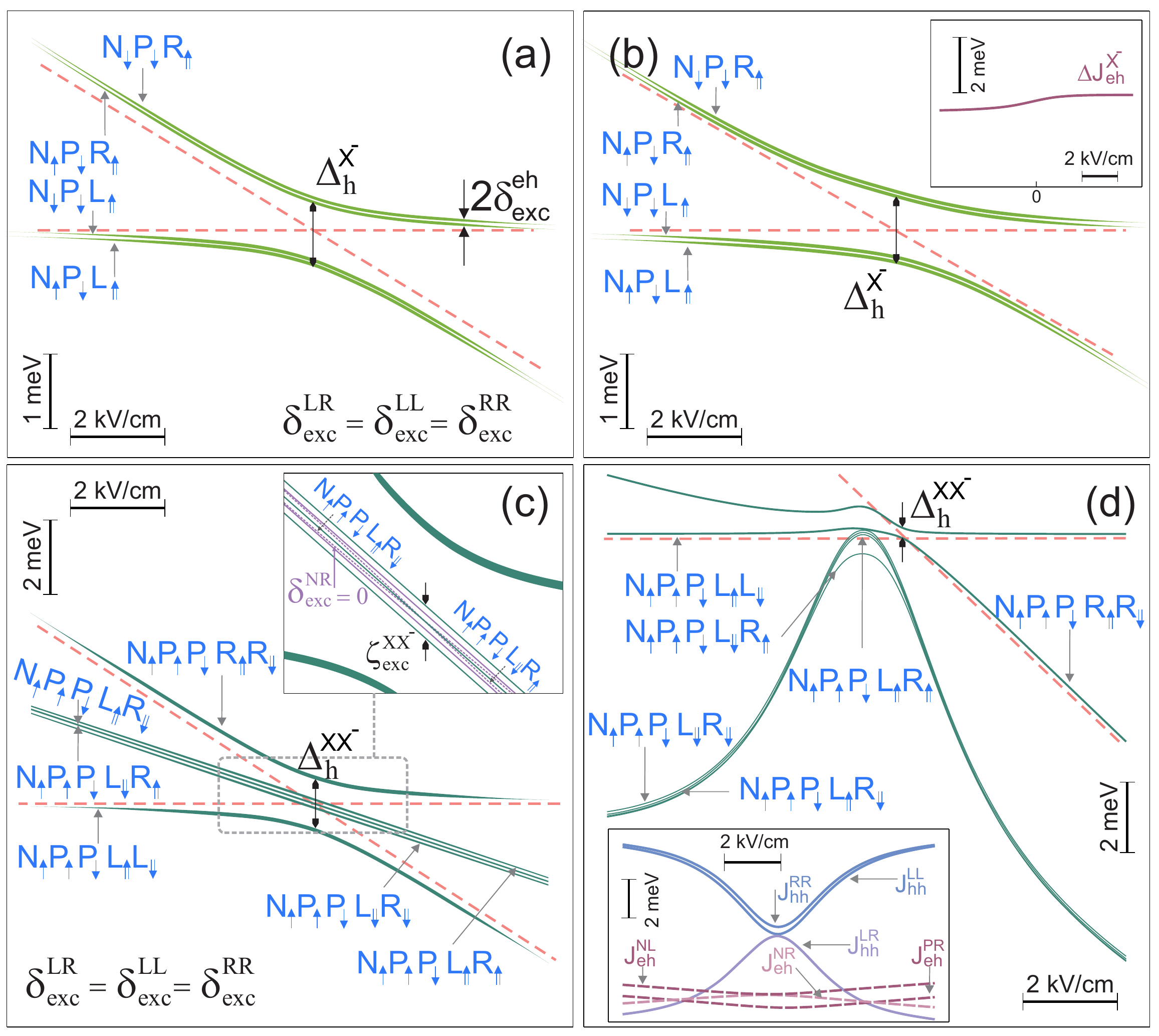}\caption{(a) Energy level diagram of the irreducible trion basis set plotted with respect to the reference energy $\zeta_{L}^{X^{-}}$ as a function of the axial electric field once the direct Coulomb interactions are switched off. 
$N_{\uparrow}P_{\downarrow}L_{\Uparrow}$
and 
$N_{\uparrow}P_{\downarrow}R_{\Uparrow}$
($N_{\downarrow}P_{\downarrow}L_{\Uparrow}$
and 
$N_{\uparrow}P_{\downarrow}R_{\Uparrow}$)
are tunnel coupled, $\Delta_{h}^{X^{-}}\simeq$ 892 $\mu$eV, $2\delta_{\mathrm{exc}}^{eh}=$ 122 $\mu$eV. (b) $\Delta J^{X^{-}}\neq0$ (inset), but the spin fine structure exhibits a similar pattern to (a) because the attractive interactions are comparable and evolve smoothly
as a function of the electric field. (c) Energy state diagram of negative
biexciton basis set calculated with respect to $\zeta_{L}^{XX^{-}}$
having the Coulomb interactions discarded. $\Delta_{h}^{XX^{-}}\simeq$
1.24 meV, $N_{\uparrow}P_{\uparrow}P_{\downarrow}L_{\Downarrow}R_{\Downarrow}$
and $N_{\uparrow}P_{\uparrow}P_{\downarrow}L_{\Uparrow}R_{\Uparrow}$
are each separated from the $N_{\uparrow}P_{\uparrow}P_{\downarrow}L_{\Downarrow}R_{\Uparrow}$-$N_{\uparrow}P_{\uparrow}P_{\downarrow}L_{\Uparrow}R_{\Downarrow}$
doublet by $2\delta_{\mathrm{exc}}^{eh}$ ($\zeta_{\mathrm{exc}}^{XX^{-}}=4\delta_{\mathrm{exc}}^{eh}$).
The inset illustrates the situation where the $\delta_{\mathrm{exc}}^{NR}$ and $\delta_{\mathrm{exc}}^{PL}$ interactions are assumed to be weak. The violet lines show the new energies
of the quadruplet with $N_{\uparrow}P_{\uparrow}P_{\downarrow}L_{\Downarrow}R_{\Uparrow}$ and $N_{\uparrow}P_{\uparrow}P_{\downarrow}L_{\Uparrow}R_{\Downarrow}$ crossing. (d) Coulomb interactions are switched on, $\Delta J_{1}^{XX^{-}}\neq\zeta_{L}^{XX^{-}}$ and $\Delta J_{2}^{XX^{-}}\neq\zeta_{L}^{XX^{-}}$. Noticeable change in $\Delta J_{1}^{XX^{-}}\simeq J_{hh}^{LR}-J_{hh}^{LL}$ decouples the quadruplet from the $N_{\uparrow}P_{\uparrow}P_{\downarrow}L_{\Uparrow}L_{\Downarrow}$
and $N_{\uparrow}P_{\uparrow}P_{\downarrow}R_{\Uparrow}R_{\Downarrow}$ singlets. Inset: evolution of attractive and repulsive interactions as a function of the axial electric field.}

\label{fig:XXN_XN_pureEnergies}
\end{figure}

We notice that the bright states $N_{\uparrow}P_{\downarrow}L_{\Uparrow}$
($\equiv N_{\downarrow}P_{\uparrow}L_{\Uparrow}\equiv N_{\uparrow}P_{\downarrow}L_{\Downarrow}\equiv N_{\downarrow}P_{\uparrow}L_{\Downarrow}$) and $N_{\uparrow}P_{\downarrow}R_{\Uparrow}$ ($\equiv N_{\downarrow}P_{\uparrow}R_{\Uparrow}\equiv N_{\uparrow}P_{\downarrow}R_{\Downarrow}\equiv N_{\downarrow}P_{\uparrow}R_{\Downarrow}$) are fourfold degenerate, while $N_{\downarrow}P_{\downarrow}L_{\Uparrow}$ ($\equiv N_{\uparrow}P_{\uparrow}L_{\Downarrow}$) and $N_{\downarrow}P_{\downarrow}R_{\Uparrow}$ ($\equiv N_{\uparrow}P_{\uparrow}R_{\Downarrow}$) are twofold degenerate
as long as $\delta_{\mathrm{exc}}^{NL}=\delta_{\mathrm{exc}}^{PR}=\delta_{\mathrm{exc}}^{NR}=\delta_{\mathrm{exc}}^{PL}$
and the trivial anisotropic exchange interactions are discarded. These
assumptions are not necessarily valid in a strained QDM as could be
revealed by our CI calculations, but facilitate providing a
semi-quantitative description of the exchange couplings. The total
energy spectrum of $X^{-}_{eeh=NPh}$ as a function of the axial
electric field is plotted in Figure~\ref{fig:XXN_XN_pureEnergies}(a), where the direct Coulomb interactions are artificially discarded $\Delta J_{1}^{X^{-}}=0$, and (b) where the states are Coulomb correlated $\Delta J_{1}^{X^{-}}\neq0$. The quadruplets labeled by $N_{\uparrow}P_{\downarrow}L_{\Uparrow}$ and $N_{\uparrow}P_{\downarrow}R_{\Uparrow}$ (analogous to
the doublets labeled by $N_{\downarrow}P_{\downarrow}L_{\Uparrow}$
and $N_{\downarrow}P_{\downarrow}R_{\Uparrow}$) evolve to
resonance where the anticrossing caused by the hole tunneling appears,
$\Delta_{h}^{X^{-}}=2\sqrt{t_{h}^{2}+(2\delta_{\mathrm{exc}}^{eh})^{2}}\simeq$ 892 $\mu$eV having $2\delta_{\mathrm{exc}}^{eh}=122\:\mu$eV and
$t_{h}=430\mu$eV. Relatively trivial variations of $\Delta J_{1}^{X^{-}}$,
shown in the inset of Figure~\ref{fig:XXN_XN_pureEnergies}(b), is a consequence of electron delocalization and comparable direct and indirect $e$-$h$ interactions, thus $\Delta_{h}^{X^{-}}$ is not noticeably renormalized. 

Negative biexciton includes two hole particles, thus the spin-conserved
tunneling of the hole is restricted by the Pauli exclusion principle.
This spin blockade leads to six $XX_{eeehh=N_{\uparrow}P_{\uparrow}P_{\downarrow}hh}^{-}$ doublets considering the twofold spin degeneracy of each electron. In order to resolve the corresponding energy state diagram, we consider the basis states 
$N_{\uparrow}P_{\uparrow}P_{\downarrow}L_{\Downarrow}R_{\Uparrow}$
($\equiv N_{\downarrow}P_{\uparrow}P_{\downarrow}L_{\Downarrow}R_{\Uparrow}$),
$N_{\uparrow}P_{\uparrow}P_{\downarrow}L_{\Uparrow}R_{\Downarrow}$
($\equiv N_{\downarrow}P_{\uparrow}P_{\downarrow}L_{\Uparrow}R_{\Downarrow}$),
$N_{\uparrow}P_{\uparrow}P_{\downarrow}L_{\Downarrow}R_{\Downarrow}$
($\equiv N_{\downarrow}P_{\uparrow}P_{\downarrow}L_{\Uparrow}R_{\Uparrow}$),
$N_{\uparrow}P_{\uparrow}P_{\downarrow}L_{\Uparrow}R_{\Uparrow}$
($\equiv N_{\downarrow}P_{\uparrow}P_{\downarrow}L_{\Downarrow}R_{\Downarrow}$),
$N_{\uparrow}P_{\uparrow}P_{\downarrow}R_{\Uparrow}R_{\Downarrow}$
($\equiv N_{\downarrow}P_{\uparrow}P_{\downarrow}R_{\Uparrow}R_{\Downarrow}$)
and 
$N_{\uparrow}P_{\uparrow}P_{\downarrow}L_{\Uparrow}L_{\Downarrow}$
($\equiv N_{\downarrow}P_{\uparrow}P_{\downarrow}L_{\Uparrow}L_{\Downarrow}$), then construct the $6\times6$ few-body hamiltonian $\mathcal{H}_{\mathrm{\mathit{XX^{-}}}}^{\mathrm{B}}$ as

\begin{equation}
\left(\begin{array}{cccccc}
\Delta J_{1}^{XX^{-}}-\zeta_{E_{z}} & 0 & 0 & 0 & -t_{h} & -t_{h}\\
0 & \Delta J_{1}^{XX^{-}}-\zeta_{E_{z}} & 0 & 0 & -t_{h} & -t_{h}\\
0 & 0 & \Delta J_{1}^{XX^{-}}-\zeta_{E_{z}}+2\delta_{\mathrm{exc}} & 0 & 0 & 0\\
0 & 0 & 0 & \Delta J_{1}^{XX^{-}}-\zeta_{E_{z}}-2\delta_{\mathrm{exc}} & 0 & 0\\
-t_{h} & -t_{h} & 0 & 0 & \Delta J_{2}^{XX^{-}}-2\zeta_{E_{z}} & 0\\
-t_{h} & -t_{h} & 0 & 0 & 0 & \zeta_{L}^{XX^{-}}
\end{array}\right),
\end{equation}

\begin{equation}
\Delta J_{1}^{XX^{-}}=\zeta_{L}^{XX^{-}}+J_{eh}^{NL}+2J_{eh}^{PL}-J_{eh}^{NR}-2J_{eh}^{PR}+J_{hh}^{LR}-J_{hh}^{LL}
\end{equation}
and 
\begin{equation}
\Delta J_{2}^{XX^{-}}=\zeta_{L}^{XX^{-}}+2(J_{eh}^{NL}-J_{eh}^{NR}+2J_{eh}^{PL}-2J_{eh}^{PR})+J_{hh}^{RR}-J_{hh}^{LL}.
\end{equation}
where $\zeta_{L}^{XX^{-}}$ is the total energy of $N_{\uparrow}P_{\uparrow}P_{\downarrow}L_{\Uparrow}L_{\Downarrow}$. Assuming that the direct and indirect interactions are comparable,
$\Delta J_{1}^{XX^{-}}=\zeta_{L}^{XX^{-}}$ and $\Delta J_{2}^{XX^{-}}=\zeta_{L}^{XX^{-}}$, we observe the singlet states 
$N_{\uparrow}P_{\uparrow}P_{\downarrow}L_{\Uparrow}L_{\Downarrow}$
and 
$N_{\uparrow}P_{\uparrow}P_{\downarrow}R_{\Uparrow}R_{\Downarrow}$
undergoing an "anticrossing" with an energy gap equivalent to
$\Delta_{h}^{XX^{-}}=2\sqrt{2t_{h}^{2}+(2\delta_{\mathrm{exc}}^{eh})^{2}}\simeq$ 1.24 meV; see Figure~\ref{fig:XXN_XN_pureEnergies}(c). The two singlets, however, first couple to 
$N_{\uparrow}P_{\uparrow}P_{\downarrow}L_{\Downarrow}R_{\Uparrow}$
or 
$N_{\uparrow}P_{\uparrow}P_{\downarrow}L_{\Uparrow}R_{\Downarrow}$
rather than undergoing a direct anticrossing since the two-hole tunneling
is prohibited in our model. The triplet states 
$N_{\uparrow}P_{\uparrow}P_{\downarrow}L_{\Downarrow}R_{\Downarrow}$
and 
$N_{\uparrow}P_{\uparrow}P_{\downarrow}L_{\Uparrow}R_{\Uparrow}$
are separated by $\zeta_{\mathrm{exc}}^{XX^{-}}=4\delta_{\mathrm{exc}}^{eh}$
and pass through the anticrossing region without tunnel coupling to
the singlet states, because the spin state is conserved during the
tunneling. The remaining states, 
$N_{\uparrow}P_{\uparrow}P_{\downarrow}L_{\Downarrow}R_{\Uparrow}$
and 
$N_{\uparrow}P_{\uparrow}P_{\downarrow}L_{\Uparrow}R_{\Downarrow}$,
form a singlet-triplet pair in proximity to the anticrossing. Their
degeneracy seen in Figure~\ref{fig:XXN_XN_pureEnergies}(c) returns back to the assumption of all exchange interactions being equivalent. Beyond the thin barrier limit, $\delta_{\mathrm{exc}}^{NR}$ exchange interaction is negligible, thus 
$N_{\uparrow}P_{\uparrow}P_{\downarrow}L_{\Downarrow}R_{\Uparrow}$
and 
$N_{\uparrow}P_{\uparrow}P_{\downarrow}L_{\Uparrow}R_{\Downarrow}$
split by $\zeta_{\mathrm{exc}}^{XX^{-}}=2\delta_{\mathrm{exc}}^{eh}$
as illustrated in the inset of Figure~\ref{fig:XXN_XN_pureEnergies}(c).

In Figure~\ref{fig:XXN_XN_pureEnergies}(d), the energy state diagram of $XX_{eeehh=N_{\uparrow}P_{\uparrow}P_{\downarrow}hh}^{-}$
is depicted against the electric field, taking the Coulomb interactions
into account ($\Delta J_{1}^{XX^{-}}\neq\zeta_{L}^{XX^{-}}$ and $\Delta J_{2}^{XX^{-}}\neq\zeta_{L}^{XX^{-}}$).
$h$-$h$ repulsive interactions significantly change as a function
of axial localization, leading to large variations in $J_{hh}^{LL}$,
$J_{hh}^{LR}$ and $\Delta J_{1}^{XX^{-}}$ versus the electric field as shown in the inset. The Coulomb splitting drastically renormalizes
the energy levels of the quadruplet 
$N_{\uparrow}P_{\uparrow}P_{\downarrow}L_{\Downarrow}R_{\Uparrow}$,
$N_{\uparrow}P_{\uparrow}P_{\downarrow}L_{\Uparrow}R_{\Downarrow}$,
$N_{\uparrow}P_{\uparrow}P_{\downarrow}L_{\Downarrow}R_{\Downarrow}$,
$N_{\uparrow}P_{\uparrow}P_{\downarrow}L_{\Uparrow}R_{\Uparrow}$,
and decouples the hole spin singlets, 
$N_{\uparrow}P_{\uparrow}P_{\downarrow}L_{\Uparrow}L_{\Downarrow}$,
and 
$N_{\uparrow}P_{\uparrow}P_{\downarrow}R_{\Uparrow}R_{\Downarrow}$,
from 
$N_{\uparrow}P_{\uparrow}P_{\downarrow}L_{\Downarrow}R_{\Uparrow}$ 
and 
$N_{\uparrow}P_{\uparrow}P_{\downarrow}L_{\Uparrow}R_{\Downarrow}$,
in the vicinity of the anticrossing gap. The splitting can be suppressed by intermixing the dot-barrier materials and thus delocalizing the $\ket{h_{0}}$ and $\ket{h_{1}}$ molecular orbitals. Since the negative trion states are rather insensitive to Coulomb interactions, the anticrossings visible in the $N_{\uparrow}P_{\uparrow}P_{\downarrow}hh \rightarrow NPh$
transitions are mainly due to the tunnel-coupling of final states. Extending the same principles to positively charged complexes, the
spin fine structure of the positive trion $X^{+}$ should exhibit
a similar pattern as $XX^{-}$ as long as the hole tunneling is concerned.
The presence of three holes in the positive biexciton $XX^{+}$, however,
further complicates the effective hamiltonian. We do not study the
total energy of $XX^{+}$ here, but we predict observing trivial
decouplings because the $h$-$h$ repulsions effectively cancel out
each other, i.e. $J_{hh}^{LL}+2J_{hh}^{RL}-J_{hh}^{RR}-2J_{hh}^{LR}\leq2t_{h}$.

\section{Cascade Transitions under Transverse Magnetic Field}

CI method incorporates both isotropic and anisotropic $e$-$h$ exchange
interactions $\delta_{\mathrm{exc}}^{eh}=-\beta^{\mathrm{exc}}\mathbf{\boldsymbol{\sigma}_{\mathit{e}}}\cdot\mathbf{J}_{h}$,
where $\beta^{\mathrm{exc}}$ is the coupling coefficient and $\mathbf{J_{\mathit{h}}}$ is the hole angular momentum operator. $\beta^{\mathrm{exc}}$ coefficients
of mutual $e$-$h$ exchange interactions are calculated using $\bra{\psi_{\sigma_{e}}^{e}\psi_{\sigma_{h}}^{h}}\mathcal{C}\ket{\psi_{\sigma_{h}}^{h}\psi_{\sigma_{e}}^{e}}=\bra{\psi_{\sigma_{e}}^{e}}V_{\sigma_{e},\sigma_{h}}^{e,h}\ket{\psi_{\sigma_{h}}^{h}}$,
where $V_{\sigma_{e},\sigma_{h}}^{e,h}$ is the mean-field potential
caused by the mixed orbital $\Braket{\psi_{\sigma_{e}}^{e}|\psi_{\sigma_{h}}^{h}}$ \cite{Stier1999}. Upon applying a transverse (non-axial) magnetic
field in the Voigt geometry, the magnetic-field-induced coupling terms
between the bright and dark excitons appear, enhancing the energy
splittings and providing access to the desirable spin states. The
off-diagonal mixing matrix elements $\bra{\alpha_{i}}\mathcal{H}_{\mathbf{B}}^{ij}\ket{\alpha_{j}}$
between two arbitrary configurations $\alpha_{i}$ and $\alpha_{j}$
is given by $\mathcal{H}_{\mathbf{B}}^{ij}=\frac{\mu_{B}}{2}g_{eff}^{\perp}\mathbf{B}_{\perp}\cdot\mathbf{\boldsymbol{\sigma}}_{ij}$,
where $g_{eff}^{\perp}$ stands for the effective transverse g-factor
at a particular axial electric field, and $\mathbf{\boldsymbol{\sigma}}_{ij}$
is the spin flip operator associated with configurations $\alpha_{i}$
and $\alpha_{j}$. The dark-bright mixing gives rise to a nonzero
many-body oscillator strength of dark states, introducing extra optically-active decays in proximity to the unperturbed transitions. Activation of dark transitions may lead to the emergence of unwanted spectral features and several $\Lambda$ systems \cite{Xu2007}, comprising metastable electron and hole ground states with unwanted spin configurations. Therefore, they shall be eliminated from the spectral window before being subjected to any additional post-processing. 

One striking feature in the energy level diagram of QDMs, as compared
to single QDs, is the existence of available bright channels leading
to both $\ket{e_{\uparrow}}$ and $\ket{e_{\downarrow}}$ electron
final states. Hence, the requisite for activating dark states via
a transverse magnetic field is basically eliminated, and the Faraday
configuration (parallel to the strong quantization axis) could also
be employed for the spin control or tuning the charged complexes.
Our single particle calculations based on the gauge invariant discretization
method \cite{Andlauer2008} show that due to the orbital resemblance,
the hole effective g-factors are comparable for the low-lying $s$-shells,
that is $g_{\ket{s_{h,L}}}^{x}=0.262$ and $g_{\ket{s_{h,L}}}^{x}=0.258$,
and smaller than the electron effective g-factors, $g_{\ket{s_{e,N}}}^{x}=-1.08$ and $g_{\ket{s_{e,P}}}^{x}=-1.042$, at $E_{z}=-5\,\mathrm{kV/cm}$.
We comment that strong localization of hole orbitals under electric
field results in smaller g-factors and less Zeeman splitting of charged
states. The reason lies in the fact that g-factor can be explained as the measure of  deformation the orbital undergoes in response to the external magnetic field. Once the hole orbital becomes squeezed in the real space, its Fourier transform spreads out in $k$-space. The hole particle then gains further kinetic energy if its orbital reshapes. The reluctance to exchange the kinetic energy prohibits the orbital to deform by the magnetic field and the g-factor drops. Therefore, although excitons are fully entangled in the vicinity of anticrossing, they suffer from larger Zeeman shifts under the magnetic field.

\begin{figure}[H]
\centering
\includegraphics[scale=0.47]{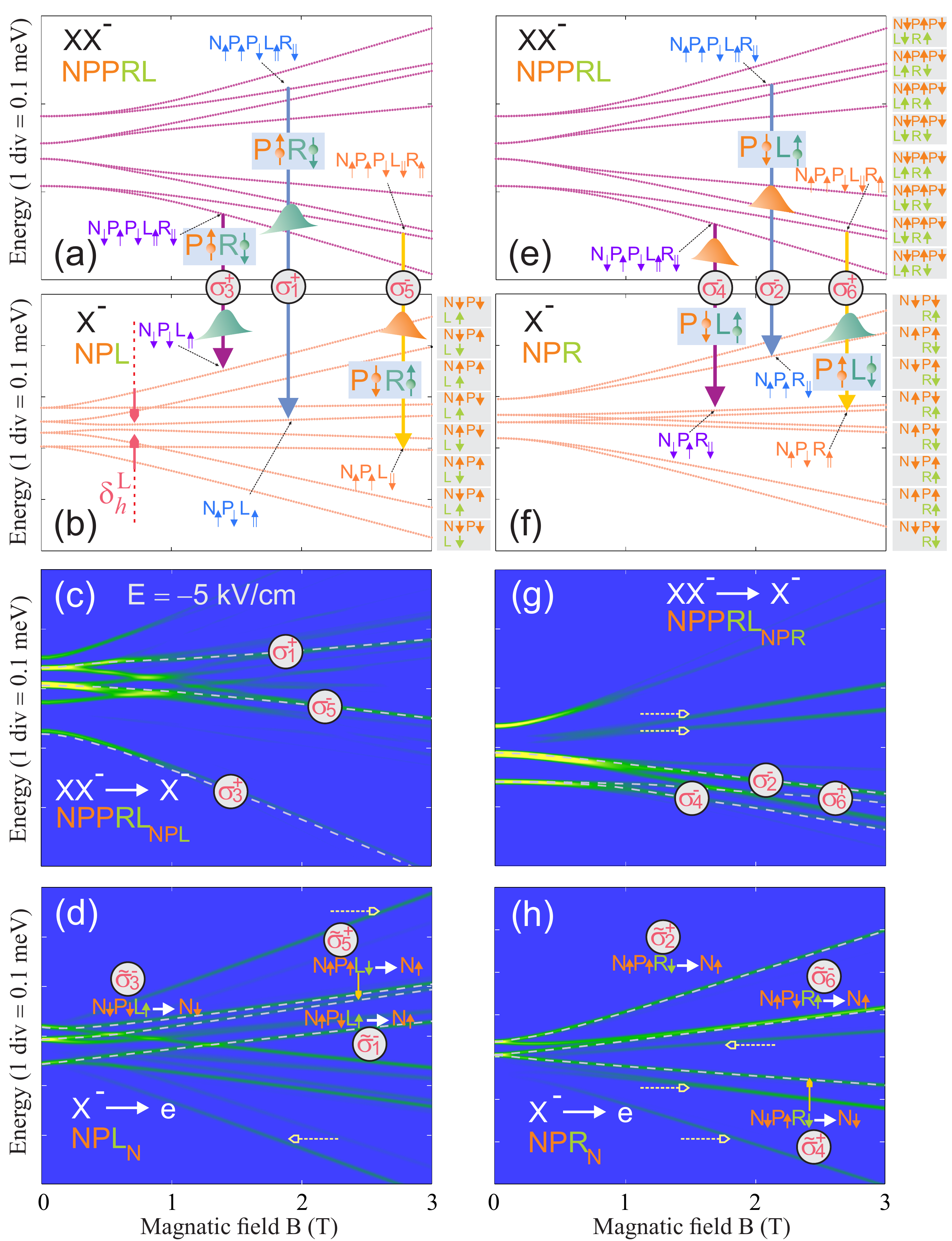}\caption{(a) and (e) Negative biexciton $NPPLR$ and, (b) and (f), negative trions, $NPL$ and $NPR$, energies versus the {[}100{]}-oriented magnetic field at $E_{z}=-5\,\mathrm{kV/cm}$. Spin configurations of $X^{-}$ and $XX^{-}$ are exchange and Zeeman coupled, and the dominant character attributed to each configuration at $\mathbf{B}_{x}=3\mathrm{T}$ is indicated at the right side of each panel. Configurations with all the spins flipped are approximately degenerate at $\mathbf{B}_{x}=0$. Intermediate trions configured in $NPL$ and $NPR$ form following $\left\langle P_{\sigma_{e}}|R_{\sigma_{h}}\right\rangle $ and $\left\langle P_{\sigma_{e}}|L_{\sigma_{h}}\right\rangle $ recombinations, respectively. Three pairs of $XX^{-}_{X^{-}}$ bright transitions, (c) and (g), exhibiting large enough oscillator strengths are selected: $\sigma_{1}^{+}$ ($\sigma_{2}^{-}$), $\sigma_{3}^{+}$ ($\sigma_{4}^{-}$) and $\sigma_{5}^{-}$ ($\sigma_{6}^{+}$). The subsequent $X{}^{-}_{e}$ transitions, (d) and (h), are labeled by $\tilde{\sigma}_{1}^{-}$ ($\tilde{\sigma}_{2}^{+}$), $\tilde{\sigma}_{3}^{-}$ ($\tilde{\sigma}_{4}^{+}$) and $\tilde{\sigma}_{5}^{+}$ ($\tilde{\sigma}_{6}^{-}$). Transitions $\sigma_{1}^{+}$ ($\tilde{\sigma}_{1}^{-}$), $\tilde{\sigma}_{2}^{+}$ ($\sigma_{2}^{-}$), $\sigma_{3}^{+}$ ($\tilde{\sigma}_{3}^{-}$) and $\tilde{\sigma}_{4}^{+}$ ($\sigma_{4}^{-}$) correspond to the polarized photons shown in Figure \ref{fig:Cascade_negativeBiexciton}. The energies of final electron states $\{N_{\uparrow},\, N_{\downarrow}\}$ are not shown here. Horizontal arrows in (c-d) and (g-h) mark dark states gaining noticeable oscillator strengths upon Zeeman mixing.}

\label{QDMunderMagneticField}
\end{figure}

Figure ~\ref{QDMunderMagneticField}(a) and (b) illustrate the energy spectrum of $XX^{-}$ and $X^{-}$ for different spin configurations in $NPPLR$ and $NPL$ components, respectively, as a function of the transverse magnetic field
$\mathbf{B}_{x}$ at $E_{z}=-5\,\mathrm{kV/cm}$. We selected three
exemplary configurations from the $XX^{-}$ energy state spectrum
comprising two bright recombinations with electrons localized in $\ket{P}$. Transitions labeled as $\sigma_{1}^{+}$, $\sigma_{3}^{+}$
and $\sigma_{5}^{-}$ correspond to the bright $\left\langle P|R\right\rangle $ recombinations. The oscillator strength of $XX^{-}\{NPPLR\}_{X^{-}\{NPL\}}$
transitions is given in Figure ~\ref{QDMunderMagneticField}(c). The anticrossings visible at low magnetic fields correspond to the coupling to dark trion features or the bright-bright trion mixing; for example, $\delta_{h}^{L}=11\,\mu$eV is the hole anticrossing energy between $N_{\downarrow}P_{\uparrow}R_{\Uparrow}$ and $N_{\downarrow}P_{\uparrow}R_{\Downarrow}$ states. Such anticrossing behaviour changes the dominant character of Zeeman-coupled spin configurations in the course of magnetic field variations. The measure of anticrossing energy pertains to the effective g-factor of the coupled spin states. The primary character of each spin configuration at $\mathbf{B}_{x}=$ 3T is indicated in Figure~\ref{QDMunderMagneticField}(a-b) and (e-f).

The asymmetric character of $s$-shell molecular orbitals reveals comparing Figures~\ref{QDMunderMagneticField}(b) and (f) where localization of the trion-bound hole is merely altered, but the effective $g$-factors of trions are different. Similar to Figure \ref{QDMunderMagneticField}(c), the oscillator strengths of the trions are shown in Figure~\ref{QDMunderMagneticField}(d) labeled by $\tilde{\sigma}_{1}^{+}$, $\tilde{\sigma}_{3}^{+}$ and $\tilde{\sigma}_{5}^{-}$. Compared to $\sigma_{i=1,3}^{+}$ ($\sigma_{5}^{-}$), the trion transitions $\tilde{\sigma}_{i=1,3}^{-}$ ($\tilde{\sigma}_{5}^{+}$) undergo relatively smaller energy change versus $\mathbf{B}_{x}$.
The maximum energy variation observed for each individual transition
however does not exceed 200 $\mu$eV under this range of magnetic
field $\mathbf{B}_{x}<3\mathrm{T}$. Right panel of Figure~\ref{QDMunderMagneticField} depicts the same set of data for $XX^{-}\{NPPLR\}_{X^{-}\{NPR\}}$ transitions: the initial $XX^{-}$ states are identical but the intermediate trion states here emerge following $\left\langle P|L\right\rangle $ recombination, also the polarization
of the emitted photon is reversed. The spectral dispersion
through coupling to multiple states is considerably suppressed in
both $XX^{-}_{X^{-}}$ and $X^{-}_{e}$ spectra as compared to the alternative path due to the less number of crossings (anticorssings) seen in the $NPR$ trion. 
\\

\section{Photon pair concurrence}

Full analysis of the spin-photon pair entanglement should be performed in an experimental framework or via a complete mathematical model incorporating the timing details and spin dynamics. An explicit parameter to incorporate the energy and linewidth of two arbitrary transitions in QDM spectrum is the photon-pair concurrence. In our particular case, the tripartite concurrence could be estimated once the spin dynamics is known \cite{Gao2006}. Here we exclusively analyze the photon indistinguishability through calculating the concurrence $C$, which is ideally $C=0$ for separable states and $C=1$ for the maximally entangled states in the regular scheme, but never reaches above 0.73 in the time reordering scheme (cross entanglement). In order to verify the entanglement between the photon pair and the third particle (electron spin), their correlations in the current spin-polarization basis as well as the rotated basis shall be measured. Previous observation
of these correlations are reported for the spin-single photon entanglement
\cite{DeGreve2012}. In the following, we narrow down our model to the cross-entanglement scheme. The findings are then extensible to the regular scheme.

According to Figure~\ref{fig:Cascade_negativeBiexciton}, the concurrence of photon pairs relies upon $\Gamma_{XX^{-}\rightarrow X_{\lambda}^{-}}$, $\Gamma_{X_{\lambda}^{-}\rightarrow e}$, $\lambda\in\{\sigma^{+},\sigma^{-}\}$, and the normalized detuning $\Delta_{\omega}^{\aleph}=[\delta^{\sigma_{1,4}^{+}}+\mathcal{O}(\delta_{\mathrm{exc}},\Delta_{\mathbf{B}})]/2\Gamma_{X_{\sigma^{+}}^{-}\rightarrow e}$, where $\Gamma_{\alpha_{i}\rightarrow\alpha_{j}}$ stands for half the spontaneous emission rate of the recombining exciton: the concurrence drops with increasing $\Gamma_{XX^{-}\rightarrow X_{\lambda}^{-}}/\Gamma_{X_{\lambda}^{-}\rightarrow e}$ and $\Delta_{\omega}^{\aleph}$ \cite{Avron2008}; see Supplemental Information. Both these parameters can be optimized via frequency conversion technique to maximize the concurrence, because the process leads to frequency and linewidth conversion of initial $XX_{X^{-}}^{-}$ and $X_{e}^{-}$ photons. Assuming that the dephasing linewidth could become as narrow as $30\,\mu\mathrm{eV}$ in defect-free nanowire-QDs \cite{Dalacu2012}, the normalized detuing drops significantly subsequent to the frequency conversion process, which is able to enhance $\Gamma$ even by one order of magnitude \cite{DeGreve2012}.

\begin{figure}
\includegraphics[scale=0.7]{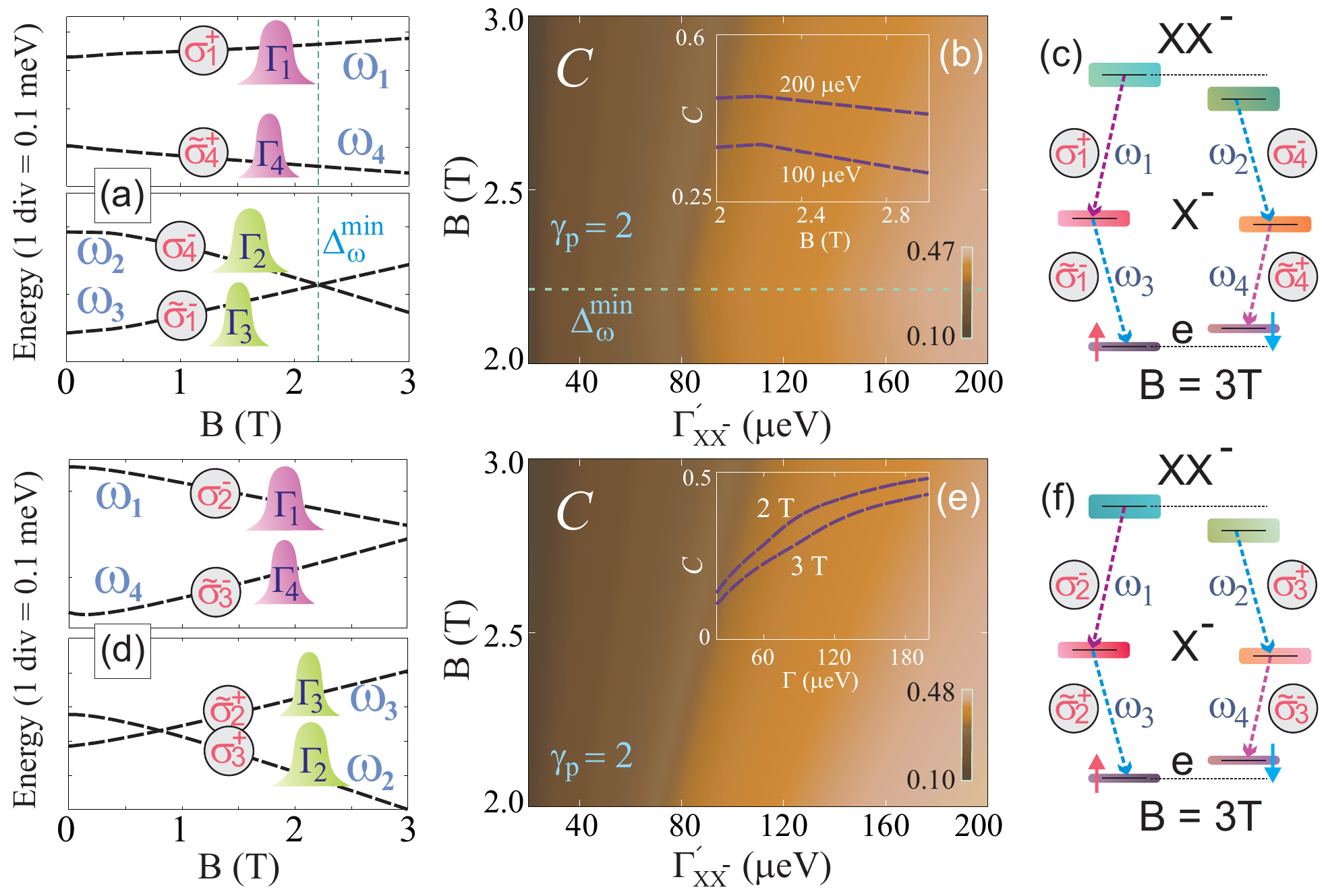}\caption{(a), (d) Energy of photon pairs created across generations in cascade paths $\mathcal{P_{\mathrm{1}}}$ and $\mathcal{P_{\mathrm{4}}}$, $\mathcal{P_{\mathrm{2}}}$ and $\mathcal{P_{\mathrm{3}}}$. The respective linewidths of photons before frequency conversion are illustrated in a sense that upper transitions possess less dephasing lifetime.
The frequency of each transition is labeled according to the $XX^{-}\rightarrow X_{\lambda}^{-}\rightarrow e$ energy diagrams shown in (c) and (f). (b) Concurrence of $\mathcal{P_{\mathrm{1}}}$-$\mathcal{P_{\mathrm{4}}}$
cascades versus magnetic field and $\Gamma_{XX^{-}\rightarrow X_{\lambda}^{-}}^{'}$;
$\gamma_{p}=2$. The inset shows concurrence as a function of $\mathbf{B}_{x}$
for $\Gamma_{XX^{-}\rightarrow X_{\lambda}^{-}}^{'}=$100 $\mu$eV and 200 $\mu$eV.
Maximum value appears at $\Delta_{\omega}^{\mathrm{min}}$ where $\omega_{2}=\omega_{3}$. (e) Concurrence of photon pairs in $\mathcal{P_{\mathrm{2}}}$-$\mathcal{P_{\mathrm{3}}}$ cascades; $\gamma_{p}=2$. Inset: concurrence versus $\Gamma_{XX^{-}\rightarrow X_{\lambda}^{-}}^{'}$ plotted for $\mathbf{B}_{x}$= 2 T and 3 T.}

\label{fig:QDMconcurrence}
\end{figure}

Above linewidths relate to the photon emission rates proportional
to $\bra{XX_{\lambda}^{-}}\mathcal{H}_{\mathrm{em}}\ket{X_{\lambda}^{-}}$
and $\bra{X{}_{\lambda}^{-}}\mathcal{H}_{\mathrm{em}}\ket{e}$. Here,
$\mathcal{H}_{\mathrm{em}}$ is the coupling hamiltonian to the optical
modes $l$, $\mathcal{H}_{\mathrm{em}}=\sum_{c;l;\lambda}g_{l\lambda}^{c}\hat{a}_{l\lambda}^{\dagger}\hat{b}_{\lambda}+\mathrm{H.c.}$
where $g_{l\lambda}^{c}$ is the oscillator strength of transition
$c=\{XX_{\lambda}^{-}\rightarrow X{}_{\lambda}^{-},X{}_{\lambda}^{-}\rightarrow e\}$,
$\hat{b}_{\sigma^{+}}=\hat{h}_{\Downarrow}\hat{c}_{\uparrow}$, $\hat{b}_{\sigma^{-}}=\hat{h}_{\Uparrow}\hat{c}_{\downarrow}$, and $\hat{a}_{l\lambda}^{\dagger}$ creates a photon in $l$th optical mode with polarization $\lambda$. Figures~\ref{fig:QDMconcurrence}(a) and (d) depict the energy evolution of photons created in paths $\mathcal{P_{\mathrm{1}}}$ and $\mathcal{P_{\mathrm{4}}}$ ($\mathcal{P_{\mathrm{2}}}$ and $\mathcal{P_{\mathrm{3}}}$) as a function of the transverse magnetic field. For the sake of simplicity, we assume that the average linewidth of upper transitions in Figure~\ref{fig:Cascade_negativeBiexciton}  after frequency conversion is twice the linewidth of lower ones: $\gamma_{p}=\Gamma_{XX^{-}\rightarrow X_{\lambda}^{-}}^{'}/\Gamma_{X_{\lambda}^{-}\rightarrow e}^{'}\approx2;$
$\Gamma_{\sigma_{1}^{+}}^{'}=\Gamma_{\sigma_{4}^{-}}^{'}=2\Gamma_{\tilde{\sigma}_{1}^{-}}^{'}=2\Gamma_{\tilde{\sigma}_{4}^{+}}^{'}$.
Although this ratio is independent of the dephasing linewidth and primarily
relies on the time resolution of the laser source pumping the PPLN, our
assumption does not pose any constraint to the generality of results.
We further assume that the frequencies of single photons are down-converted
with the same ratio, i.e. $\omega_{i=1,2,3,4}^{'}=\eta\omega_{i=1,2,3,4}$.
The concurrence of the photon pairs in paths $\mathcal{P_{\mathrm{1}}}$
and $\mathcal{P_{\mathrm{4}}}$ is then given by \cite{Avron2008,Pathak2009}

\begin{equation}
C=\frac{4}{\pi^{2}}\int\int\frac{\Gamma_{\sigma_{1}^{+}}\Gamma_{\tilde{\sigma}_{4}^{+}}W_{o}(\omega_{m},\omega_{n})}{(\omega_{m}+\omega_{n}-\Omega_{1;XX^{-}}-i\Gamma_{\sigma_{1}^{+}}^{'})(\omega_{m}+\omega_{n}-\Omega_{4;XX^{-}}+i\Gamma_{\sigma_{4}^{-}}^{'})}\times \\ \frac{d\omega_{m}d\omega_{n}}{(\omega_{m}-\omega_{3;X^{-}}^{'}-i\Gamma_{\tilde{\sigma}_{1}^{-}}^{'})(\omega_{m}-\omega_{4;X^{-}}^{'}+i\Gamma_{\tilde{\sigma}_{4}^{+}}^{'})},\label{eq:4-6}
\end{equation}
where $\Omega_{1;XX^{-}}=\omega_{1}^{'}+\omega_{3}^{'}$ and $\Omega_{4;XX^{-}}=\omega_{2}^{'}+\omega_{4}^{'}$; see Figure~\ref{fig:QDMconcurrence}(c). The simplest additional phase above $W_{o}(\omega_{m},\omega_{n})$
can be a linear phase with time delay $\tau_{o}$, that is $W_{o}(\omega_{m},\omega_{n})=\mathrm{exp}[i(\omega_{m}-\omega_{n})\tau_{o}]$.
Since the detuning between the two paths $\mathcal{P_{\mathrm{1}}}$
and $\mathcal{P_{\mathrm{4}}}$, $\Delta_{\omega}=|\omega_{1}^{'}-\omega_{4}^{'}|+|\omega_{3}^{'}-\omega_{2}^{'}|$,
is constantly nonzero in our setup, the time delay suggested by Pathak
and Hughs, $\tau_{o}=\mathrm{ln}(1+\Gamma_{XX^{-}\rightarrow X_{\lambda}^{-}}^{'}/2\Gamma_{X_{\lambda}^{-}\rightarrow e}^{'})/\Gamma_{XX^{-}\rightarrow X_{\lambda}^{-}}^{'}$, does not necessarily optimize the concurrence here. The optimum $\tau_{o}$,
however, could be resolved empirically once the actual level broadenings
are determined. 

The concurrence of photon pairs versus the magnetic
field and $\Gamma_{XX^{-}\rightarrow X_{\lambda}^{-}}^{'}$ is plotted in Figure~\ref{fig:QDMconcurrence}(b). The magnetic field magnitude is set above 2 T where the spin initialization and readout are experimentally feasible. The transition
oscillator strengths are comparable and steady in this range, permitting
the detuning $\Delta_{\omega}$ to be the only parameter restraining
the concurrence. According to the continuous increase in $\omega_{1}^{'}$-$\omega_{4}^{'}$ splitting, local extremum occurs at $\Delta_{\omega}^{\mathrm{min}}$ at which $\tilde{\sigma}_{1}^{-}$ and $\sigma_{4}^{-}$ coincide.
The inset shows how noticeably the concurrence is improved once $\Gamma_{XX^{-}\rightarrow X_{\lambda}^{-}}^{'}$ increases from 100 $\mu$eV up to 200 $\mu$eV. Figure~\ref{fig:QDMconcurrence}(e) depicts the same plot for the concurrence of photon states across generations in $\mathcal{P_{\mathrm{2}}}$ and $\mathcal{P_{\mathrm{3}}}$. Energy detunings $|\omega_{4}^{'}-\omega_{1}^{'}|$ and $|\omega_{2}^{'}-\omega_{3}^{'}|$ evolve oppositely versus the magnetic field, thus the concurrence remains insensitive towards its variations, see inset Figure~\ref{fig:QDMconcurrence}(e).

Above example demonstrates that level broadenings up to half the photon-photon
detuning energy ($\sim200\,\mu$eV) could improve the cascade concurrence
considerably. On the other hand, exerting the magnetic field in the Faraday configuration would keep the dark states inactive and conserve the bright transition oscillator strengths undispersed by prohibiting any spin-flip mixing. Coherent control of the spin rotation  may then be more favorable in Faraday configuration owing to its clean emission spectrum. We note that intermediate exchange-coupled trion
states are assumed to be immune to the single spin flip, e.g. $N_{\uparrow}P_{\downarrow}L_{\Uparrow}\,\rightarrow N_{\downarrow}P_{\downarrow}L_{\Uparrow}$,
or the cross-dephasing process, e.g. $N_{\uparrow}P_{\downarrow}L_{\Uparrow}\,\rightarrow N_{\downarrow}P_{\uparrow}L_{\Uparrow}$,
during the excitation cycle. Principally, weak axial quantization of molecular orbitals gives rise to smaller spin relaxation times of the exciton-bound electron and hole \cite{Tsitsishvili2003,Tsitsishvili2005}. The hole molecular orbitals are squeezed at higher electric field here, but the electron orbitals preserve a fixed volume regardless of the electric field magnitude. 

Apart from the photons across generations in Figure~\ref{fig:Cascade_negativeBiexciton}, entanglement between the photons within generations could also be established via the frequency conversion technique. The first emitted photons in paths $\mathcal{P_{\mathrm{1}}}$ and $\mathcal{P_{\mathrm{3}}}$ (or second emitted photons in $\mathcal{P_{\mathrm{2}}}$ and $\mathcal{P_{\mathrm{4}}}$) are in fact energetically separated only by the measure of exchange and Zeeman splittings. This implies that the charged biexciton cascades offer several path options to produce tripartite entangled states. In addition, we notice that the level of concurrence calculated here delicately depends on $\gamma_{p}$. Without applying the frequency conversion technique, $\gamma_{p}$ represents a relative parameter comparing the natural linewidths of $XX^{-}\rightarrow X_{\lambda}^{-}$ and $X_{\lambda}^{-}\rightarrow e$ transitions. Therefore, the concurrence is essentially untouched and determined merely by the QDM structure. Upon a frequency conversion, each $\Gamma_{XX^{-}\rightarrow X_{\lambda}^{-}}^{'}$ and $\Gamma_{X_{\lambda}^{-}\rightarrow e}$ could be manipulated individually, leading to a tunable range of $\gamma_{p}$. we predict that higher values of concurrence are obtainable by tailoring the $XX^{-}$ and $X^{-}$ level broadenings with different scales as concurrence grows monotonously once $\Gamma_{XX^{-}\rightarrow X_{\lambda}^{-}}^{'}/\Gamma_{X_{\lambda}^{-}\rightarrow e}^{'}$ decreases \cite{Avron2008}. 

Finally, we comment on how a QDM could be exploited to generate GHZ or W entangled states of photons. The existence of two low-lying $s$-shells in the energy level spectrum of QDM leaves room for three correlated excitons to form a triexciton $XXX$ under proper pumping conditions. Sequential decays of $XXX$ down to the ground state, $XXX\rightarrow XX \rightarrow X\rightarrow G$, create three correlated photons whose color indistinguishability could be manipulated via the frequency conversion technique. Different configurations of initial and intermediate bright states then prepare a set of decay paths comprising diverse combinations of photon polarizations $\ket{\sigma^{\pm}_{1}\sigma^{\pm}_{2}\sigma^{\pm}_{3}}$. A two-path combination might be able to produce a GHZ state, such as $1/\sqrt{2}(\ket{\sigma^{+}_{1}\sigma^{+}_{2}\sigma^{+}_{3}}+\ket{\sigma^{-}_{1}\sigma^{-}_{2}\sigma^{-}_{3}})$, while a three-path combination might give rise to a W-state. We notice that the color distinguishability in a spin-free state merely counts on direct and exchange energies as no external magnetic field is involved. Provided that the combinations at each level, $XXX$ or $XX$ or $X$, are chosen from spin-flipped configurations, the color distinguishability pertains only to exchange energies which are typically limited as compared to the Zeeman splitting.

\section{Summary}

We investigated the feasibility of producing tripartite spin-photon
pair entanglement from charged biexcitons in quantum dot molecules.
In the proposed structure, two quantum dots are coupled sharing their
$s$-shell ground states with a relatively small hybridization energy
of the hole particle ($t_{h}<0.5$ meV) that facilitates switching between
different configurations without extensive cost of tunneling energies.
We particularly analyzed the case of In(Ga)As stacked QDs embedded
in {[}001{]}-oriented GaAs nanowires. However, the results can be
generalized to other types of QDMs sustaining strongly correlated
orbitals. We showed that the weak quantization of electron opens new
channels of recombination for higher order entanglement. Our calculations
in the few-body framework revealed that charged biexcitons and excitons
exhibit dominant spectral features as a direct consequence of existing
mutual interactions. The variety of available transitions then narrows
down to a few favorable initial and final states. Neglecting the small
exchange splittings,  $XX^{-}\rightarrow X^{-}\rightarrow e$ cascades
represent relative energy matching under axial electric filed, while
$XX^{+}\rightarrow X^{+}\rightarrow h$ channels lack any color coincidence
under our range of fields. This is primarily attributed to the large
electron hybridization energy. By providing an exemplary cascade of
$XX^{-}\rightarrow X^{-}\rightarrow e$ transitions, we examined the
double dot spin fine structure and Zeeman shifts under the magnetic
field in the Voigt geometry. The strong quantization of the hole particle
away from anticrossing lead to its small transverse g-factor, thus
charged complexes undergo less Zeeman shift. Upon empirically approved
range of magnetic field required for the spin manipulation ($\mathbf{B}<3\mathrm{T}$),
photon energy detunings were calculated below $400\,\mu$eV. Sufficient
photon concurrence is then achievable by manipulating the level broadenings
through the frequency conversion technique. Our analysis is a demonstration
of how properly-sorted multi-particle states could be exploited to
create multi-partite entangled states in an engineerable solid state
source.

\section{Acknowledgment}

This research is supported by NSERC Discovery Grant, Waterloo Institute
for Nanotechnology and Institute for Quantum Computing at the University
of Waterloo. M. Khoshnegar thanks G. Bester at the Max Planck Institute
for Solid State Research and Christopher Haapamaki at Coherent Spintronics
group in IQC for their fruitful discussions. We acknowledge IST services
at University of Waterloo for supporting the computational facilities.

\bibliographystyle{h-physrev3}
\bibliography{ref}

\end{document}